\documentclass[preprint]{aastex6}
\usepackage{mathrsfs}
\usepackage{amssymb,amsmath,float}
\usepackage{rotating}
\usepackage{color}

\newcommand\var{{\rm var}}
\newcommand\cov{{\rm cov}}

\newcommand\cT{\mathcal{T}}

\newcommand\cI{\mathcal{I}}

\newcommand\frs{|f_{rs}|^2}

\newcommand\al{\alpha}

\newcommand\de{\delta}
\newcommand\ep{\epsilon}

\newcommand\si{\sigma}
\newcommand\ta{\tau}

\newcommand\vp{\varphi}

\newcommand\om{\omega}

\newcommand\De{\Delta}

\newcommand\Om{\Omega}

\newcommand\<{\langle}
\renewcommand\>{\rangle}

\newcommand\ie{\emph{i.e.}}
\newcommand\eg{\emph{e.g.}}

\newcommand\beq{\begin{equation}}
\newcommand\eeq{\end{equation}}
\newcommand\bea{\begin{eqnarray}}
\newcommand\eea{\end{eqnarray}}
\newcommand\bal{\begin{align}}
\newcommand\eal{\end{align}}

\newcommand\fr{\frac}

\newcommand\ba{\mathbf{a}}
\newcommand\bb{\mathbf{b}}

\renewcommand\bal{\mbox{\mathbfmath$\alpha$}}

\begin{document}

\title{A METHOD OF ENHANCING THE DETECTION SENSITIVITY OF TRANSIENT SOURCES IN TIME SERIES WITH GAUSSIAN STATIONARY NOISE}

\author{Richard Lieu$^1$ and Kristen A. Lackeos$^2$}

\affil{$^1$Department of Physics, University of Alabama,
Huntsville, AL 35899\\}

\affil{$^2$ NASA Postdoctoral Program Fellow, NASA Marshall Space Flight Center, Huntsville, AL 35812, USA}

\begin{abstract}
The Gaussian phase noise of intensity time series is demonstrated to be drastically reduced when the raw voltage data are digitally filtered through an arbitrarily large number $n$ of orthornormal bandpass profiles (eigen-filters) sharing the same intensity bandwidth, and the resulting intensity series are co-added.  Specifically, the relative noise variance of the summed series at the resolution of one coherence time or less, goes down with increasing $n$ as $1/n$, although (consistent with the radiometer equation) the advantage gradually disappears when the series is bin averaged to lower resolution.  Thus the algorithm is designed to enhance the sensitivity of detecting transients that are smoothed out by time averaging and too faint to be visible in the noisy unaveraged time series, as demonstrated by the simulation of a weak embedded time varying signal of either a periodic nature or a fast and unrepeated pulse.  The algorithm is then applied to a 10 minute observation of the pulsar PSR 1937+21 by the VLA, where the theoretical predictions were verified by the data. Moreover, it is shown that microstructures within the time profile are better defined as the number $n$ of filters used increases, and a periodic signal of period $1.86 \times 10^{-5}$~s ($53.9$~kHz) is discovered in the pulse profile. Lastly, we apply the algorithm to the first binary black hole merger detected by LIGO, GW150914. We find the SNR of the mean peak intensity increases as $\sqrt{n}$ and cross correlation of the event between the LIGO-Hanford-Livingston detector pair increases with filter order $n$.

\end{abstract}

\section{Introduction}

Communication in the radio differs from most other wavelengths of the electromagnetic spectrum in one key respect: the number of photons in some observed dataset is sufficiently large to consign the Poisson fluctuation in the arrival time of each photon (also called shot noise) to the realm of insignificance relative to another inevitable noise source, {\it viz.} Gaussian noise (also called photon bunching noise) which is a characteristic attribute of the wave nature of light.  In quantitative terms, the relative importance of these two fundamental noise components in naturally occurring light is the photon occupation number $n_0$, which in the context of a time series of measured intensities is the number of arriving photons per unit frequency bandwidth $\De\nu$ per unit coherence time $\ta \approx 1/\De\nu$.

The criterion on $n_0$ is derived in many textbooks and review articles on the subject such as \cite{lou00}.  Explicitly, as we shall see below, for phase noise to be more important than shot noise the inequality \beq \fr{n_0 T}{\ta} \gg 1,  \label{classical} \eeq where $T$ is the sampling time, must be satisfied.  Physically it means the average number of photons arriving within a sampling interval is $\gg 1$.  If there are $N$ time contiguous samplings that comprise a total exposure of $\cT = NT \gg \ta$, the ratio of square of the mean intensity to the noise power will for Gaussian chaotic light be of order the number of coherence time intervals within the total exposure, {\it viz.}~$\approx \cT/\ta$.
This result is often referred to as the radiometer equation (\cite{bur10,chr85}), and stems from the simple realization that in the intensity time series the noise fluctuation tends to randomly cancel among the various coherence segments but not within each of them.  Another argument would interpret $\cT/\ta \approx \cT\De\nu$ as the number of independent Fourier modes in the intensity time series, \ie~the accuracy of a mean intensity measurement over the interval $\cT$ is not controlled by the number $N$ of sampling intervals $T$ but the product of the two.  Thus $\cT/\ta$ may also be interpreted as the minimum number of samples required to fully determine the characteristics of all the constituent modes -- it is not necessary to sample the intensity (or voltages) more than the Nyquist limit of once per coherence time $\ta$.  The radiometer equation was recently shown by \cite{nai15} to be a fundamental unsurpassable limit on the uncertainty in the mean intensity of Gaussian noise dominated light over an exposure time $\cT \gg \ta$.

The purpose of this paper is to demonstrate, both theoretically and with real data, a method to enhance the detection of transient signals embedded in Gaussian noise without violating  the radiometer equation.  We emphasize that our proposed methodology is not in conflict with the radiometer equation, nor does it produce a biased estimate of the mean intensity of Gaussian chaotic light.  Rather, it merely lowers the noise variance from some original value $\si_I^2$ to a smaller value $\si_I^{'2} < \si_I^2$, by stretching the correlation length of the intensity time series.  It shall be shown that the mean intensity over an interval $\cT \gg \ta$ is still governed by the radiometer equation, but an embedded source of duration $\De t$ satisfying $\De t \gtrsim \ta$\footnote{The fastest transient allowed by the Fourier bandwidth theorem is $\De t \approx\ta$} is detected more readily by the new approach.
Specifically, the present paper is about a new algorithm which 
 enhances Gaussian noise-limited variations in the data existing on timescales 
 on the order of, but not less than, the coherence time of the intensity time series.

The plan of the paper is to begin with a revisit of the basics of Gaussian noise, including a derivation of the radiometer equation.  This will be followed by a mathematical treatment of the way our proposed methodology of noise suppression enhances the signal-to-noise on the scale of $\ta$, whilst maintaining consistency with the radiometer equation. The methodology is translated into an algorithm which is applied to two types of simulated signals embedded
in Gaussian noise.  

The algorithm is applicable to any Gaussian noise limited signal, 
where the variations one is seeking to enhance last the order of the coherence time or longer, $\gtrsim \tau$. 
In signals with periods $\gg \tau$,  real periodic or quasi-periodic modulations on smaller 
timescales therein are shown to be enhanced.  
We apply the algorithm to a VLA observation of the millisecond pulsar PSR B1937+21 where 
it will be shown that, depending upon the number of eigen-filters being used in the data processing stage, 
features of the pulsar light curve within a narrow band becomes increasingly more resolved. 
From the ACF of the pulse profile the statistical significance of the correlation is shown to increase 
with filter order. Additionally, a hidden periodicity at higher harmonic number is revealed. 
The algorithm is lastly applied to Laser Interferometer Gravitational Wave Observatory (LIGO) event 
GW150914, the first gravitational-wave (GW) detection, and the analysis is found to be in agreement theory.  

The types of LIGO sources most amenable to this type of analysis will be those existing on timescales 
lasting the order of the coherence time or longer, although there is an optimum duration beyond which the 
signal-to-noise ratio gradually drops back to the conventional value given by the radiometer equation. 
The optimum duration is longer the larger the number of eigenfilters employed. 
One can adjust the coherence time of the observed radiation by 
choosing various filter bandwidths, for example as we do here to produce oversampled data for the radio analysis.
 The only caveat is that one should be careful to construct an eigenfilter which is symmetric across the chosen 
 analysis bandwidth.

%The method presented is applicable in general to any Gaussian noise limited signal existing on timescales 
%lasting the order of the coherence time or longer. 
In our analysis and proof-of-principle demonstrations we focus on enhancing astrophysical signals, 
specifically applying the technique to a known pulsar and GW signal. 
In a GW search pipeline, for unknown signals, short-duration instrumental and 
environmental noises masquerading as astorphysical signals are still an issue. 
Having two or more detector streams responding to a common signal becomes essential to 
vetoing this type of noise in LIGO.  Given this constraint, our algorithm is 
best suited for a GW detection pipeline that can also accomodate the cross correlation of $n$-part, 
co-added intensity data from multiple detectors. Likewise, procedures such as whitening and radio frequency interference 
removal are required, before an `eigenfilter search' for astrophysical (or artificial, extaterrestrial) signals is conducted.

The limitation of the algorithm is that the advantage gradually disappears when the timescale of the transient 
becomes $\gg \tau$. In the case of pulsar signal, quasi-periodic modulations which exist within one pulse period
but exceed timescales $\gg \tau$ will not be enhanced. In Section 4 the method's applicability is further explored.

\section{Two-point correlation of voltage and intensity of stationary Gaussian light}

Owing to the approximation of high occupation number, $n_0 \gg 1$, shot noise is negligible and one can calculate $n$-point amplitude correlations classically by treating the quantum operators as c-numbers, \cite{wan89}.  For a Gaussian chaotic light (also known as Gaussian thermal light), such as especially radio noise, the voltage may be written as a linear superposition of Fourier amplitudes at random phase, {\it viz.}~\beq V(t) = \sum_{j=1}^{N_m} a_j~e^{i(\om_j t + \phi_j)}, \label{Vt} \eeq where $a_j = a(\om_j)$, the number of modes is $N_m = \cT\De\nu$ with $\cT$ being the total exposure time and $\De\nu$ the bandwidth, and the phases $\phi_j$ are random and uncorrelated.

The time series of $V(t)$ is evidently governed by a stationary stochastic process with vanishing ensemble mean, $\<V(t)\> =0$.  The two point function of $V(t)$ is \beq \<V(t) V^* (t+t')\> =\left< \sum_j a_j~e^{i(\om_j t + \phi_j)} \sum_k a^*_k~e^{-i[\om_k (t+t')+ \phi_k]}\right>, \label{VtVt'} \eeq where the range of the summation for $j$ and $k$ are now dropped, with the understanding that it is from $1$ to $N_m$ for both.  The ensemble average of the two point function, $\<V(t) V^* (t+t')\>$ is obtained by noting that in \eqref{VtVt'} unless $j=k$ this average vanishes.  Thus the result is independent of $t$, and (moreover) one may replace $t'$ by $t$ by writing \beq \<V(0) V^* (t)\> = \sum_j |a_j|^2 e^{-i\om_j t}. \label{VV'} \eeq  In the continuum limit this becomes the Fourier transform of the amplitude, \ie~the Fourier transform is centered at $\om=\om_0$ and spans the bandwidth $\De\om = 2\pi\De\nu$, $\<V(0) V^* (t)\>$ would typically be finite over some time interval of size $\ta\approx 1/\De\nu$, which may be taken as the coherence length of the voltage autocorrelation function.

At zero lag $t=0$, $\<V(0) V^* (t)\>$ becomes the mean intensity.  The intensity two-point function is \bea I(t)I(t+t') &=& V(t)V^*(t)V(t+t')V^*(t+t') \notag\\
&=& \sum_j a_j~e^{i(\om_j t + \phi_j)} \sum_k a^*_k~e^{-i(\om_k t + \phi_k)}
\sum_p a_p~e^{i[\om_p (t+t') + \phi_p]} \sum_q a^*_q~e^{-i[\om_q (t+t') + \phi_q]}. \notag\\\eea  As before, due to the random phases the ensemble average $I(t)I(t+t')$ is obtained by noting the only two combinations of summation indices that yield finite contributions, {\it viz.}~$j=k,~p=q$ and $j=q,k=p$.  Thus again, independently of $t$ so that one can set $t=0$ and rewrite $t'$ as $t$,
\bea \<I(0)I(t)\> - \<I\>^2  &=& \<V(0)V^*(0)V(t)V^*(t)\> - \<V(0) V^* (0)\>^2 \notag\\
&=& \left| \sum_j a_j a_j^* e^{-i\om_j t} \right|^2. \label{I0It}
\eea
Thus in the continuum limit the intensity covariance function is the modulus square of the Fourier transform of $|a (\om)|^2$.  If, as in the case of the voltage correlation, $a(\om)$ spans the bandwidth $\De\om$, the intensity covariance function will extend to the coherence length $\ta$ which is the reciprocal of the bandwidth.

To be very precise about the relationship between coherence length and bandwidth, let the spectrum of an arriving radiation be of the form \beq |a(\om)|^2 = \fr{n_0}{\sqrt{2\pi}\cT} e^{-(\om - \om_0)^2\ta^2/2}. \label{asq} \eeq  This spectrum may be intrinsic to the source itself, or due to a bandpass filter being physically or digitally applied to an otherwise spectrally flat source.  Then the summation in \eqref{I0It}, when evaluated as an integral, becomes \beq \<I(0)I(t)\> - \<I\>^2 = \left|\cT \int_{-\infty}^\infty~|a(\om)|^2 e^{-i\om t}~d\om \right|^2 = \fr{n_0^2}{\ta^2}~e^{-t^2/\ta^2}. \label{GI0It} \eeq  Since, from \eqref{VV'}, the mean intensity is $\<V(0)V^*(0)\> = n_0/\ta$, \eqref{GI0It} yields the normalized covariance function (autocorrelation function, ACF) \beq  \fr{\<I(0)I(t)\> - \<I\>^2}{\<I\>^2} = e^{-t^2/\ta^2}, \label{normcov} \eeq from which one reads off the relative variance of the intensity as \beq \fr{\si_I^2}{\bar{I}^2} = \fr{\<I(0)I(0)\> - \<I\>^2}{\<I\>^2} = 1. \label{relvar} \eeq  Note that \eqref{relvar} is a fundamental property of Gaussian noise, {\it viz.} the variance equals the square of the mean intensity. More generally, for any spectrum $|a(\om)|^2$, \eqref{normcov} may be written as \beq  \fr{\<I(0)I(t)\> - \<I\>^2}{\<I\>^2} = |f(t)|^2, \label{gnormcov} \eeq where \beq f(t) =  \fr{\int_{-\infty}^{\infty} |a(\om)|^2 e^{-i\om t} d\om}{\int_{-\infty}^{\infty} |a(\om)|^2 d\om}. \label{ft} \eeq is a complex function satisfying $f(t) =1$ at $t=0$ while having finite values in the range $|t| \lesssim \ta \approx 1/\De\nu$. Also from \eqref{asq} and the relation $\cT\de\nu = 1$ where $\de\nu$ is the mode spacing, \beq \cT\int_{-\infty}^{\infty}\fr{1}{2\pi} |a(\om)|^2 d\om = \sum_j |a(\om_j)|^2 = \fr{n_0}{\ta}, \label{meanI} \eeq
is the ensemble mean intensity (or flux).

Let us next examine what happens when the intensity measured over some small but finite interval $T \ll \ta$.  The result may be expressed as \beq I_r = \fr{1}{T} \int_{t_r-T}^{t_r} dt'\, I(t'). \label{IT0} \eeq The normalized covariance between two intensity measurements which took place during intervals $T$ centered at times $t_r$ and $t_s$ is, from \eqref{gnormcov}, \beq \fr{{\rm cov} (I_r, I_s)}{\<I\>^2} = \frs, \label{fjk1} \eeq where \beq \frs = |f(t_r - t_s)|^2 = |f((r-s)T)|^2 \label{frs} \eeq also has the properties $\frs =1$ for $r=s$ and $\frs = 0$ for $|t_r - t_s| \gg \ta$.

Turning to the variance of these measurements, it is given by \beq \fr{{\rm var} (I_r)}{\<I\>^2} = \fr{{\rm cov} (I_r, I_s)}{\<I\>^2} = 1. \label{varIj} \eeq  Next we see the effect when many of these small samples are time contiguously averaged to form the mean intensity \beq I_\cT = \fr{1}{N} \sum_{r=1}^N I_r \label{IT1}, \eeq over the much longer interval $\cT = NT \gg \ta$ (because $N \gg 1$).  In this case the variance of $I_\cT$ is \beq \var(I_\cT) = \fr{1}{N^2}\sum_{r,s=1}^{N} \cov(I_r,I_s). \label{vc} \eeq
When we substitute from \eqref{fjk1}, we can convert the sum over $u=r-s$ to the Gaussian integral
\beq \sum_u |f(uT)|^2 \approx \fr{1}{T}\int dt\,|f(t)|^2 \approx \fr{\ta}{T}, \eeq while the other one of the double sum becomes $N$.  In this way we obtain \beq \fr{\var(I_\cT)}{\<I\>^2} \approx \fr{\ta}{NT} = \fr{\ta}{\cT}, \label{radio} \eeq which is the radiometer equation.  For the Gaussian spectrum of \eqref{asq} where $|f(t)|^2$ is given by $\exp (-t^2/\ta^2)$, the right side \eqref{radio} assumes the more precise expression $\sqrt{\pi}\ta/\cT$.

It is also possible to directly calculate the variance of an intensity sample averaged over {\it any} duration $\cT$ as
 \beq \var(I_\cT) =  \fr{n_0^2}{\ta \cT} F\left(\fr{\cT}{\ta}\right), \label{varIcT} \eeq
where
 \beq F\left(\fr{\cT}{\ta}\right) = \fr{1}{\ta \cT} \int_{-\cT}^\cT dt\,(\cT-|t|)|f(t)|^2 \label{F} \eeq with
$f(t)$ being defined in \eqref{gnormcov}.  If $\cT$ is short compared to the coherence time $\ta$, $|f(t)|^2 \approx 1$ and \eqref{F} reduces to $F(\cT/\ta) \approx \cT/\ta$.  Since the ensemble mean intensity is $n_0/\ta$ from \eqref{meanI}, we arrive once again at \eqref{varIj}, {\it viz.} a relative intensity variance of unity.  On the other hand, if $\cT \gg\ta$, we must then use the limiting value of $F(x) \approx 1$ for $x \gg 1$ to arrive at $\ta/\cT$ as the approximate expression of the relative variance; for the Gaussian spectrum with $|f(t)|^2 = \exp (-t^2/\ta^2)$, this becomes $\sqrt{\pi}\ta/\cT$ as before.

\section{Gaussian noise suppression on short timescales by digital filtering}

Let there be a raw data set consisting of real voltages $V_r$ measured over times $t_r$ where $1\leq r \leq N$, $t_{r+1} - t_r =T$, and $NT = \cT$ as before.  Let the data be sampled at the Nyquist rate, {\it viz.} $T$ is of order the coherence time.  Now suppose $V_r$ is frequency filtered {\it digitally} into a much narrower bandwidth (equivalently the time series is convolved with a wide kernel) centered at the same $\om_0 = 2\pi\nu_0$ as the original mid-band frequency and with ${\al_r}$ being the resulting oversampled voltages -- oversampled because now $T$ is much less than the coherence time of the Gaussian noise fluctuations.  Suppose further that the exercise is repeated using a different filter which is also centered at $\om_0$ and much narrower than the original bandwidth, leading to the voltage series $\beta_s$ with $1 \leq s \leq N$.

We assume for simplicity that the raw spectrum is essentially flat between its lower and upper frequency limits, and the two narrow filtered intensity spectrum $|a_j|^2$ and $|b_k|^2$ (with $1 \leq j \leq N_m \approx \cT\De\nu$ and likewise for the index $k$) share the same bandwidth $\De\nu$, \ie~the filters only differ by their shapes in amplitude space $a_j$ and $b_k$.  Thus the two intensity time series $I_r = |\al_r|^2$ and $I'_r=|\beta_r|^2$ not only span the same time interval $\cT$, but also share the same radiation frequencies.  We shall illustrate these abstract notions with a concrete example below.  Following the arguments of the previous section, $I_r$ and $I'_r$ have ensemble intensities \beq \<I\> = \sum_j |a_j|^2;~{\rm and}~\<I'\> = \sum_k |b_k|^2. \label{I1I2} \eeq  The relative variance for $I_r$ is \beq \fr{\si_{I}^2}{\bar{I}^2} = \fr{\<I^2\> - \<I\>^2}{\<I\>^2} = 1, \label{varI} \eeq and the equation also holds for $I'_r$.

We now turn to the series $\cI_r = I_r+I'_r$ which has the ensemble mean of \beq \<\cI\> = \sum_j (|a_j|^2 + |b_j|^2) = \<I\> + \<I'\>, \label{Isum} \eeq and variance \beq \<\cI^2\> - \<\cI\>^2 = \<I\>^2 + \<I'\>^2 + 2(\<II'\> - \<I\>\<I'\>).   \label{Isumvar} \eeq  Of interest here are the last two terms, to be analyzed next.

Starting with \bea I(t)I'(t+t') &=& V(t)V'^*(t)V(t+t')V'^*(t+t') \notag\\
&=& \sum_j a_j~e^{i(\om_j t + \phi_j)} \sum_k a^*_k~e^{-i(\om_k t + \phi_k)}
\sum_p b_p~e^{i[\om_p (t+t') + \phi_p]} \sum_q b^*_q~e^{-i[\om_q (t+t') + \phi_q]}, \label{4ptV} \notag\\\eea one sees that when computing the ensemble average of $I(t)I'(t+t')$ at $t'=0$ the $j=k$ and $p=q$ pairings yield $\<I\>\<I'\>$, while the $j=q$ and $k=p$ pairings yield $|\sum_j a_j b^*_j|^2$ which satisfies the Schwarz inequality \beq \left|\sum_j a_j b^*_j \right|^2 < \sum_j |a_j|^2 \sum_k |b_k|^2 = \<I\>\<I'\>, \label{Sch} \eeq and the $<$ sign is always valid provided $a_j$ and $b_k$ are linearly independent sequences.  In this case, \eqref{Isumvar} and \eqref{4ptV} imply \beq \fr{{\rm var} (\cI_r)}{\<\cI\>^2} < 1, \label{varcI} \eeq where ${\rm var} (\cI_r) = \sigma_{\cI}^2 = \<\cI^2\> - \<\cI\>^2$.  

\begin{figure}
\begin{center}
\includegraphics[width=6in]{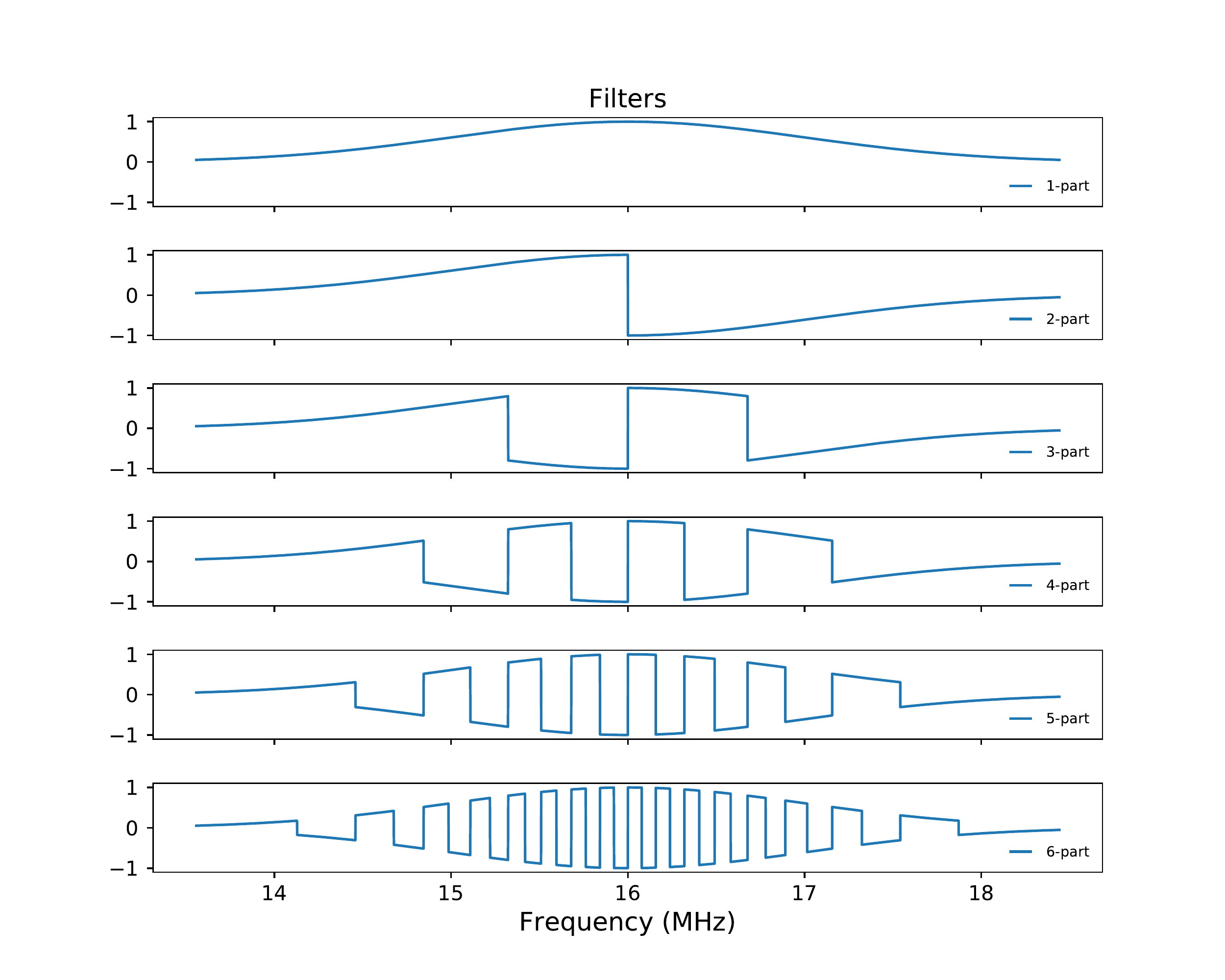}
\end{center}
\caption{The 6-part combination of digital amplitude filters to reduce the relative variance of Gaussian radio noise by $6$ times. All the structures share equal underlying areas (absolute values); and the square of each filter equals the square of the Gaussian filter at the top \ie~the intensity bandpass of the filters are the same.}
\label{filter6}
\end{figure}

Thus, we constructed a time series with an unbiased estimate of the mean (in the sense that the arithmetic mean of $\cI_r$ tends to the ensemble mean $\<\cI\>$ as the sample size tends to infinity), but with {\it less} relative variance than the standard Gaussian noise of \eqref{varIj}.  Under the particular scenario of \beq \sum_j |a_j|^2 = \sum_j |b_j|^2;~{\rm and}~\sum_j a_j b^*_j = 0, \label{zero} \eeq the relative variance of \eqref{varcI} has $0.5$ as its expectation value, which is half the standard value of \eqref{varIj}. Moreover, the technique may readily be extended to accommodate more than two digital filters.  Thus, for three filters satisfying \beq \sum_j |a_j|^2 = \sum_j |b_j|^2 = \sum_j |c_j|^2;~{\rm and}~\sum_j a_j b^*_j = 0,~\sum_j a_j c^*_j =0,~\sum_j b_j c^*_j =0 \label{mzero} \eeq the relative variance of $\cI_r = I_r + I'_r + I^{''}_r$ has the expectation value of $1/3$.  And it is also not difficult to prove the general result for $n$ filters is a relative variance of $1/n$, which can be negligibly small for arbitrarily large $n$ whilst maintaining the status of sample mean of $\cI$ as an unbiased $\<I\>$ estimator.
In Figure \ref{filter6} we illustrate the design of a set of $6$ filters (\ie~ up to $n=6$, or `6-part').

Can two filters be digitally designed to satisfy \eqref{zero} for any incident radiation having a flat spectrum within the (narrow) bandpass, where each radiation mode is taken to be of unit strength?  Assuming the filter amplitudes $a_j$ are real and positive numbers distributed symmetrically about the band center $\nu_j = \nu_0$, one viable option is to digitally set $b_j = a_j$ for $\nu_j < \nu_0$ and $b_j = -a_j$ for $\nu_j \geq \nu_0$, with $a_j$ being a real number.

As a concrete example, consider the digital filters \beq a_j =e^{-(\om_j-\om_0)^2 \ta^2/4}~{\rm for~all}~\om_j;~b_j = \pm e^{-(\om_j-\om_0)^2 \ta^2/4}~{\rm for}~\om_j \lessgtr \om_0. \label{filters} \eeq  By observing from \eqref{Isumvar} that \beq  \<\cI_r \cI_s\> - \<\cI\>^2 = \left|\sum_j a_j a^*_j e^{-i\om_j T(r-s)}\right|^2 + \left|\sum_j b_j b^*_j e^{-i\om_j T(r-s)}\right|^2 + 2\left|\sum_j a_j b^*_j e^{-i\om_j T(r-s)}\right|^2, \label{covhalf} \eeq and following the steps outlined after \eqref{4ptV}, one obtains \beq \fr{{\rm cov} (\cI_r,\cI_s)}{\<\cI\>^2} = \fr{\<\cI_r \cI_s\> - \<\cI\>^2}{\<\cI\>^2} = \fr{1}{2} e^{-(r-s)^2 T^2/\ta^2} \left\{ 1+ \left[{\rm erfi}\left(\fr{(r-s)T}{\sqrt{2}\ta}\right)\right]^2 \right\} \label{Dawson} \approx \fr{1}{2}e^{-(r-s)^2 T^2/(4\tau^2)} \eeq  By putting $r=s$ to form the variance of $\cI_r$, one readily deduces \beq \fr{\si_\cI^2}{\bar\cI^2} = \fr{{\rm var} (\cI_r)}{\<\cI\>^2} = \fr{1}{2} \label{half} \eeq since ${\rm erfi} (0)=0$.  This is consistent with the specific manifestation of \eqref{varcI} in the appropriate limit of \eqref{zero}, and indicates that the $\cI_r$ time series has less Gaussian noise on short timescales.

\begin{figure}
\begin{center}
\includegraphics[width=6in]{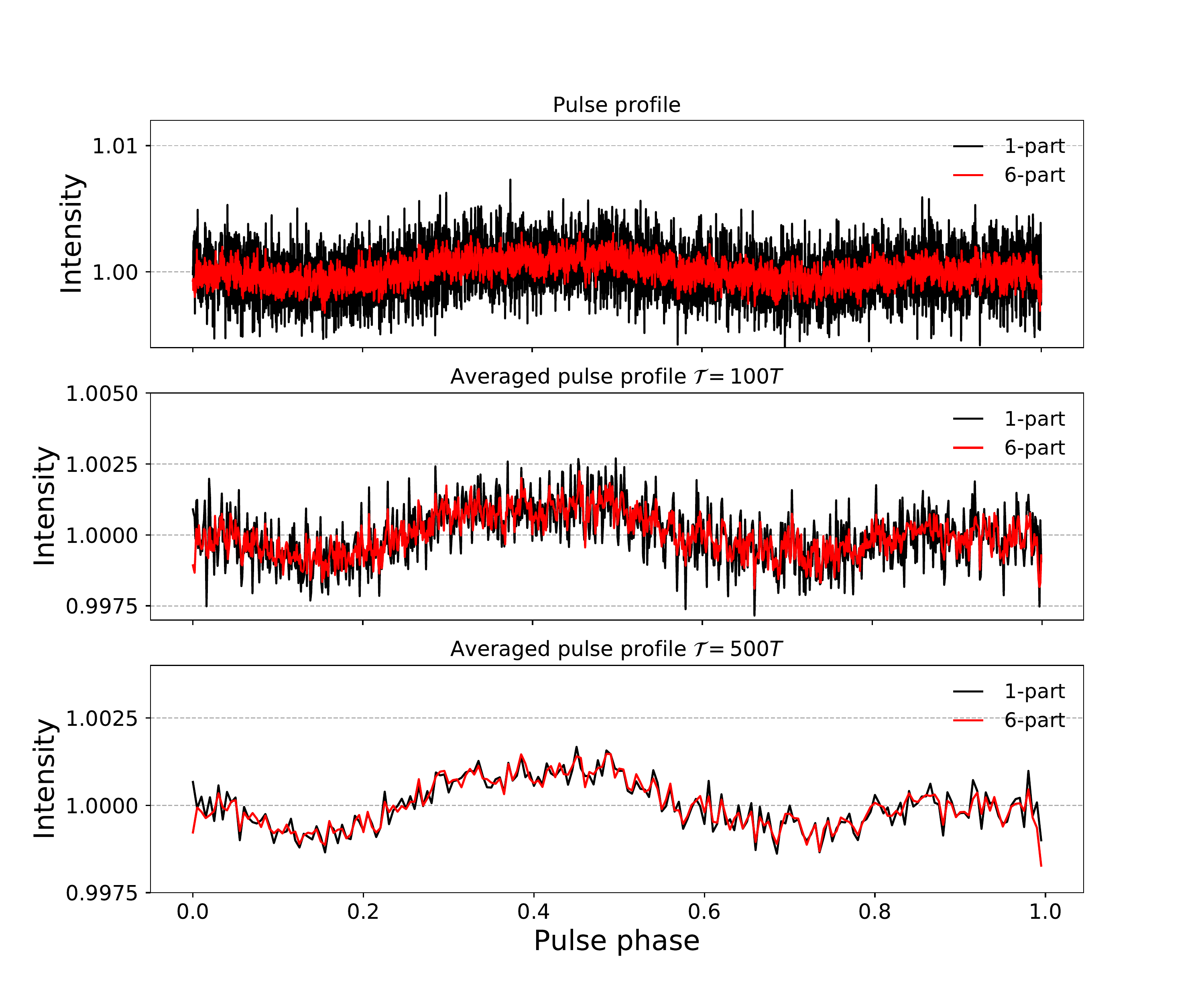}
\end{center}
\caption{The pulse profile of the entire 10 minute VLA observation of PSR 1937+21, as obtained by folding the intensity time series modulo the pulse period.  Note the considerably higher signal-to-noise of the 6-part time series relative to the conventional 1-part, when the full time resolution is applied.  The advantage goes away with averaging over larger time bins, eventually agreeing with the radiometer equation.}
\label{pulse_pro}
\end{figure}

\begin{figure}
\begin{center}
\includegraphics[width=5.5in]{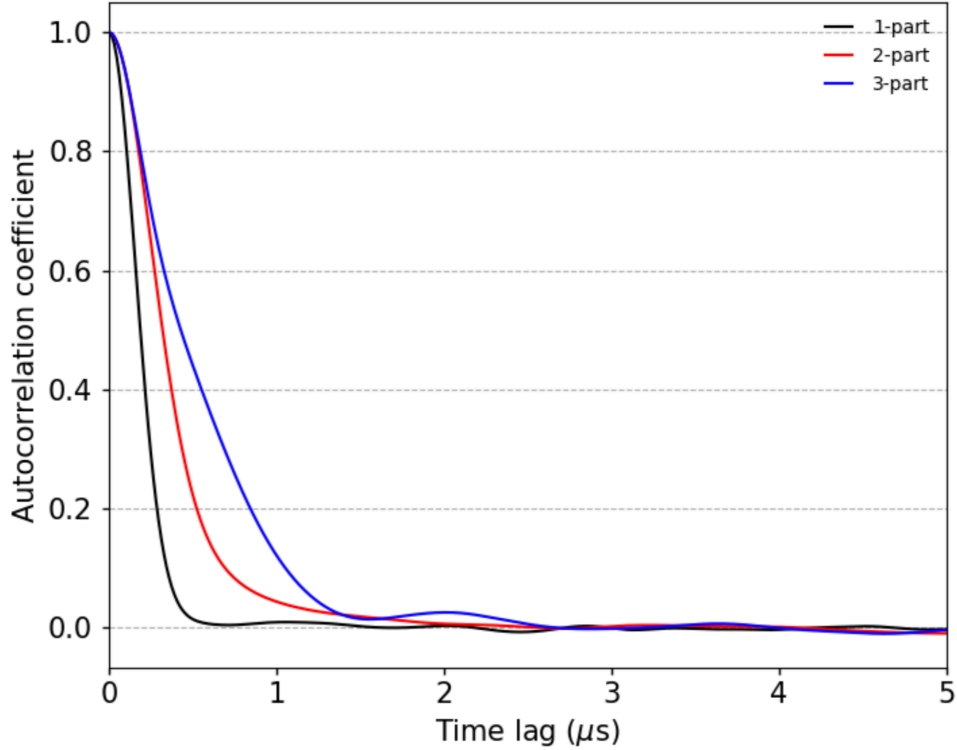}
\end{center}
\caption{Autocorrelation function (ACF) of the intensity time series of various filter combinations. The $n$-part ACF is wider than the 1-part by the factor n, and is roughly represented by a Gaussian form ${\rm exp}[-t^2/(n^2\tau^2)]$. The ACFs of 1-part through 3-part are shown here, with 1-part having the smallest correlation length and 3-part the largest.}

\label{acf}
\end{figure}

When evaluating the variance of the sample average of $\cI_r$ over a much longer timescale such as $\cT = NT \gg \ta$, however, one must apply \eqref{Dawson} to \eqref{vc}.  The result is still the radiometer equation \eqref{radio}, which confirms the claim of \cite{nai15} that the sensitivity limit imposed by this equation cannot be surpassed.  To see how enforcement of \eqref{radio} is brought about in detail, observe that although the ${\rm erfi}$ function in \eqref{Dawson} vanishes at zero lag $t=0$, it rises to a peak of order unity at $t\approx\ta$ before decaying away.  Consequently the intensity two point function in \eqref{Dawson} has a height of $0.5$ but a width $\approx 2\ta$, to be compared to the height of unity and width $\approx\ta$ for $I_r$.  Thus the {\it area} under the two point function, which gives the relative variance of flux averages over time intervals $\gg \ta$, is in fact the same for both $\cI_\cT$ and $I_\cT$.   The proof of this conclusion is extended in the Appendix to the case involving any pair of filters $\{a_j\}$ and $\{b_j\}$.  More generally, if $n$ filters are employed, the relative variance of $\cI_r$ will be reduced to $1/n$ as already explained, but the width of the covariance will be increased to $\approx n\ta$.  Thus the area under the covariance curve remains unchanged w.r.t. to its value for $I_r$, and so the relative variance of $\cI_\cT$ is still given by the radiometer equation.  Such behavior is quantitatively shown in Table 1 and Figure \ref{pulse_pro} in respect of processing a real observational data set, about which a graph of the two point function of $I_r$ and $\cI_r$ is to be found in Figure \ref{acf}.

\begin{table}[ht]

%\caption{ }

\centering

\resizebox{12cm}{!} {

\begin{tabular}{|c|c|c|c|}

\hline

 Order of processing  & ${\rm var} (\cI_\cT)/\<I\>^2$ & ${\rm var} (\cI_\cT)/\<I\>^2$ & ${\rm var} (\cI_\cT)/\<I\>^2$ \\
 & $\cT=T$ & $\cT=100T$ & $\cT=500T$ \\

\hline

1-part & 0.99  & 0.24 & 0.052  \\ \hline
2-part  & 0.50 & 0.20  &  0.049 \\ \hline

3-part  &  0.36  & 0.18 &  0.047\\  \hline

4-part &  0.27 & 0.14 & 0.044 \\ \hline

5-part  & 0.21 & 0.11 & 0.040 \\ \hline

6-part  & 0.17 & 0.08 & 0.03 \\

\hline

\end{tabular}}

\label{signaltonoise0}

\caption{The signal-to-noise ratio of the VLA intensity time series constructed from voltage data processed by filter combinations of increasing complexity, followed by intensity bin averaging.  Note the advantage of using an $n$-part filter combination with large $n$ disappears when one performs long time averaging.  Thus the methodology is only useful for the detection of transient signals that require high time resolution (hence a small averaging window $\cT$).}

\end{table}

\section{Signal-to-noise enhancement of time variable sources}

The interesting question arises, nonetheless, on what happens when a weak transient Gaussian noise source lasting a duration $\gtrsim \ta$ is embedded in the time series.
In general, one lets the amplitude of the source be enveloped by some function $\ep (t)$ which is centered at $t=t_0$ and of width $\gtrsim \ta$.  The bandwidth of frequencies generated by such an envelope is obviously within the range spanned by $\{a_j\}$ and $\{b_k\}$ (\ie~the function $\ep (t)~e^{i\om t}$ need not be resolved into modes for the analysis below).  The intensity $I(t)$ becomes \beq I(t) = V(t)V^*(t) =  \sum_j a_j~\left[e^{i(\om_j t + \phi_j)} + \ep (t)~e^{i(\om_j t + \vp_j)}\right]
\sum_k a_k^*~\left[e^{-i(\om_k t + \phi_k)} + \ep^*(t)~e^{-i(\om_k t + \vp_k)}\right], \eeq with a similar expression for $I' (t)$.  The ensemble averages are \beq \<I (t)\> = \sum_j |a_j|^2 [1+|\ep (t)|^2];~\<I' (t)\> = \sum_k |b_k|^2 [1+|\ep (t)|^2], \label{sc} \eeq and exhibit a small excess at around $t=t_0$.  Repeating the calculation of the intensity autocorrelation in the same manner as \eqref{4ptV}, noting that the two sets of random phases $\{\phi_j\}$ and $\{\vp_k\}$ are independent sets, we find as before ${\rm var}~(I_r)/\<I\>^2 = {\rm var}~(I'_r)/\<I'\>^2 =1$ to order $\ep^2$, with $I_r$ as defined in \eqref{IT0} and $\<I\>$ and $\<I'\>$ as given by \eqref{sc}, but ${\rm var}~(\cI_r)/\<\cI\>^2 < 1$.  In the special choice of filters as prescribed by \eqref{zero}, the limit of 50 \% noise reduction, {\it viz.}~\eqref{half}, is once again attained.  Since the ensemble average is $\<\cI_r\> = \<\cI (t_r)\> = \<I (t_r)\> + \<I' (t_r)\>$ for any choice of filters, where $\<I (t)\>$ and $\<I' (t)\>$ are once again as in \eqref{sc}, one sees that the noise is reduced without compromising any signal strength.

The key advantage in terms of source detection, however, is that for a transient source one does not detect it by measuring and computing the average intensity over some long duration $\cT \gg \ta$ (when the noise-to-signal ratio has returned to the limit given by the conventional radiometer equation); rather, one must enlist the full timing resolution of the observation.  More precisely, a short segment of the time series of duration $\gtrsim \ta$ is optimal to the search for such transients, and because the noise is significantly reduced with respect to the signal over the time span of $2\ta$, this offers a means of finding burst sources more sensitively without resorting to a narrower band filter to increase the coherence time $\ta$.  If more than two filters sharing the same intensity bandwidth (as the original single filter scenario) are used in accordance with the prescription of \eqref{mzero} and what followed, one would be able to look for a broader class of transients lasting the duration $\De t \lesssim n\ta$ where $n$ is the number of filters used, with a reduction of the relative variance by the factor of $n$.  The source strength, on the other hand, stays the same, so that the sensitivity of source detection is then enhanced by the factor $\sqrt{n}$.  
%The performance improvement is essentially without any limit, provided one is prepared to enlist any number of digital filters to process the raw time series of voltages.

\subsection{A periodic embedded signal}

To formulate the above in more precise terms, we consider the specific scenario of a chaotic light source with a flat spectrum (in some relevant frequency range) and emitting periodically.  The voltage at the receiver is given by \beq \ep (t) = \ep_0\sin\Om t\sum_j e^{i(\om_j t + \vp_j)}. \label{sine} \eeq Let the signal be embedded in a likewise spectrally flat background noise, so that the observed voltage at time $t$ has the normalized form \beq {\cal V} (t) = \ep(t) + \sum_k  e^{i(\om_k t + \phi_k)}. \label{bkd} \eeq  Upon Fourier transforming to the frequency domain, one obtains \beq \tilde{{\cal V}} (\om) = \fr{\pi\ep_0}{i} \sum_j e^{i\vp_j} [\de (\om_j+\Om -\om) - \de (\om_j - \Om -\om)] + 2\pi\sum_k e^{i\phi_k} \de (\om_k -\om), \label{Vsine} \eeq where $\ep_0 \ll 1$ and the phases $\{\vp_j\}$ and $\{\phi_k\}$ are uncorrelated.

After applying a narrow band filter $\{\ba\}$ to select a smaller range of frequencies, the voltage spectral amplitude becomes \beq \tilde{V} (\om) = \fr{\pi\ep_0}{i} \sum_j a_j e^{i\vp_j} [\de (\om_j+\Om -\om) - \de (\om_j - \Om -\om)] + 2\pi\sum_k a_k e^{i\phi_k} \de (\om_k -\om), \label{Vome} \eeq  Inversion back to the time domain yields \beq V(t) = \fr{\ep_0}{2i} \sum_j e^{i(\om_j t + \vp_j)} \left[ a(\om_j + \Om) e^{i\Om t} - a(\om_j - \Om) e^{-i\Om t} \right] + \sum_k a_k e^{i(\om_k t + \phi_k)} \label{Voft} \eeq  Provided $\Om \ll \om_j$ for all $\om_j$ within the selected band, one may assume $a(\om_j + \Om) \approx a(\om_j - \Om) \approx a(\om_j) = a_j$ (or equivalently $\Om$ is a small fraction of the filter bandwidth $\De\om$), in which case \eqref{Voft} simplifies to  \beq V(t) = \ep_0\sin\Om t \sum_j a_j e^{i(\om_j t + \vp_j)} + \sum_k a_k e^{i(\om_k t + \phi_k)} \label{VT} \eeq  This is the voltage time series an observer measures.

From \eqref{VT} we proceed to calculate the mean intensity in a manner analogous to \eqref{VtVt'} as the voltage two-point function at zero lag, {\it viz.} \beq \<I(t)\> = (1+\ep_0^2 \sin^2 \Om t) \sum_j a_j^2.   \label{Iex} \eeq The intensity two-point function is \beq \<I(t) I(t+\ta)\> - \<I(t)\> \<I(t+\ta)\>  =  \left|\sum_j a_j^2 e^{-i\om_j \ta} \right|^2 [1+ \ep_0^4 \sin^2 \Om t\sin^2 \Om (t+\ta)] \approx \left|\sum_j a_j^2 e^{-i\om_j \ta} \right|^2, \label{varex} \eeq where the approximation sign means one discarded a term of order $\ep_0^4 \lll 1$ (or the square of the ratio of the signal intensity to the background intensity) relative to the term kept. This indicates that the variance is much more dominated by the background noise than the mean intensity.

\subsection{Detection sensitivity}

For two intensity time series ensuing from orthonormal filters $\{\ba\}$ and $\{\bb\}$ satisfying \eqref{zero}, the intensity two-point function yields \beq \<I_1 (t) I_2 (t+\ta)\> - \<I_1 (t)\>\<I_2 (t+\ta)\> = \left|\sum_j a_j b_j e^{-i\om_j \ta} \right|^2 [1+ \ep_0^4 \sin^2 \Om t\sin^2 \Om (t+\ta)] \approx \left|\sum_j a_j b_j e^{-i\om_j \ta} \right|^2 \label{covex} \eeq and vanishes exactly when $\ta=0$.  Thus, when the two are combined to form the total intensity $\cI = I_1+I_2$, the signal of interest is manifested as an intensity difference (modulation) between two times $t_r$ and $t_s$, {\it viz.} \beq \<\cI_r - \cI_s\> = 2\ep_0^2 \ba\cdot\ba (\sin^2 \Om t_r - \sin^2 \Om t_s ), \label{sdiff} \eeq where it is assumed that $\ba\cdot\ba = \bb\cdot\bb$ in accordance with the normalizing condition of the filters as given by the first part of \eqref{zero}.  The noise variance, on the other hand, is (by \eqref{covex} with $\ta =0$) $\<\cI^2\> - \<\cI\>^2 = 2(\ba\cdot\ba)^2$ if $|t_r - t_s| \gg 2/\De\om$ (or $n/\De\om$ in the case of $n$ orthonormal filters), and $\<\cI^2\> - \<\cI\>^2 \approx 0$ if $|t_r - t_s|$ is otherwise.  Thus the signal-to-noise ratio is $\sqrt2\ep_0^2 \eta$ where $\eta = \sin^2 \Om t_r - \sin^2 \Om t_s$ , or $2\ep_0^4 \eta^2$ if expressed as the square of the signal intensity divided by the noise variance (as is the case when one estimates the signal significance in terms of the power spectral density at frequency $2\Om$).  Likewise, it can readily be shown that if only one intensity time series $I_1$ (or $I_2$) is employed the corresponding signal-to-noise ratios are $\ep_0^2 \eta$ (or $\ep_0^4 \eta^2$)  but if the intensities of $n$ orthogonally filtered voltage time streams are added they become $n \ep_0^2 \eta/\sqrt{n}$ (or $n\ep_0^4 \eta^2$) .  In this way the advantage of using multiple orthogonally filtered voltage streams is indicated.

It should also be emphasized that although on one hand the proposed technique works optimally when $\Om$ is much smaller than the filter bandwidth $\De\om = 2\pi \De\nu$ as explained after \eqref{Voft}, one does not require $n\Om$ to satisfy the same criterion.  In fact, if $n$ orthogonally filtered voltage streams are enlisted the optimal frequency $\Om$ of the embedded signal is \beq \Om \ll \De\om~{\rm but}~ n\Om \gtrsim \De\om. \label{crite} \eeq This is because when $n\Om < \De\om$ one can average the intensity data over $n$ coherence lengths (\ie~the timescale $n/\De\om$) without smoothing out the signal oscillation, in which case the relative variance of the summed intensity of the $n$ co-added intensity time series approaches the single intensity series scenario, as noted in Section 3, and the signal-to-noise advantage of the former over the latter is no longer remarkable.  The overall message is that for a small value of $\Om$, more orthogonally filtered voltage time series have to be combined to achieve significant detection sensitivity of the oscillating signal.

\subsection{Power spectrum of intensity variation}

To further elaborate upon the above analysis,  we calculate the power spectrum of intensity fluctuation, \ie~the modulus square of the Fourier transform of the intensity time series, which has the expectation value \beq \<|\tilde\cI(\om)|^2\> = \int_{-\cT/2}^{\cT/2} dt_2 \int_{-\cT/2}^{\cT/2} dt_1 e^{i\om (t_2 - t_1)} \<\cI(t_1) \cI(t_2)\> . \label{powspec} \eeq  Now as long the phases $\{\vp_j\}$ and $\{\phi_j\}$ in \eqref{Vsine} are uncorrelated, use can be made of equations like \eqref{varex} and \eqref{covex} to deduce that \bea \<\cI (t_1) \cI(t_2)\> &=& \<\cI_b\>^2 (1+\ep_0^2\sin^2 \Om t_1 +\ep_0^2\sin^2 \Om t_2 + \ep_0^4 \sin^2 \Om t_1 \sin^2 \Om t_2) \notag\\
& & +~\xi (t_2 - t_1) (\ep_0^4 \sin^2 \Om t_1 \sin^2 \Om t_2 + 1), \label{2Ipt} \eea  where \beq \xi (t) = \<I_b (0) I_b(t)\> - \<I_b\>^2, \label{xi} \eeq is the two-point function of the background intensity (\ie~intensity $I_b$ in the absence of the periodic signal).

When \eqref{2Ipt} is substituted into \eqref{powspec}, the first three terms on the right side of \eqref{2Ipt} contribute nothing to the signal power $\<|\tilde \cI(\om)|^2\>$ in the vicinity of $\om=2\Om$, while the next three terms yield respectively the ones on the right side of the following equation \beq \<|\tilde \cI(\om)|^2\> = \<\cI_b\>^2 \left[\fr{\pi^2 \ep_0^4}{16}  \de^2 (\om-2\Om) +  \fr{\sqrt{\pi}\ep_0^4 \cT\ta}{16} e^{-n^2 \ta^2 (\om-2\Om)^2/4} + \sqrt{\pi}\cT\ta  e^{-n^2 \ta^2 \om^2/4}\right], \label{lines} \eeq where $\ta\approx 1/\De\om \ll \cT$ is the width of the $I_b$ (\ie~background intensity) two-point function under the single filter scenario\footnote{The inequality is there to ensure that the limits of integration in \eqref{powspec} are effectively $-\infty$ and $\infty$ for the $t$ integration of $\xi (t)$ and $\xi (t)\cos 2\Om t$, where $t=t_2 - t_1$}.  In arriving at \eqref{lines} the form of $\xi (t)$, for the various orthonormal filter combinations depicted in Figure \ref{filter6}, are approximated as Gaussians of width $n\ta$, {\it viz.}~\beq \xi(t) \approx \fr{\<\cI_b\>^2}{n} e^{-t^2/(n^2 \ta^2)}  \label{xih} \eeq to enable a relatively simple analytic expression for \eqref{lines}.
%The last two terms of \eqref{lines} simply reflect the fact that the Fourier transform of the last term of \eqref{2Ipt} is approximately given by transforming $\sin^2 \Om t$ (where $t=t_2 - t_1$) between the limits $-\cT/2$ and $\cT/2$, because of the temporal extent of $\xi (t)$ in \eqref{xih}.

To interpret the three terms on the right side of \eqref{lines}, the Dirac delta function in the first term is the abstract representation of a single channel spike of width $2\pi/\cT$ and height $\cT/(2\pi)$; thus the spectral amplitude of this term is $\ep_0^4 \<I_b\>^2 \cT^2 /64$ which spans the single spectral channel of width $\de\om = 2\pi/\cT$ and centered at $\om = 2\Om$ .  The second term is also due to the presence of periodic signal, it has an amplitude $\sqrt{\pi}\ep_0^4 \<I_b\>^2 \cT\ta/16$ which is smaller than the first term by the factor $\approx \cT/\ta \gg 1$, and width $\approx 1/(n\ta)$ which includes many channels because $\cT\gg n\ta$.  Evidently the first term is the tall and narrow `resonance line' at frequency $2\Om$ that one can most readily detect as symptomatic (proof) of the periodic signal.  Lastly the third term, a line centered at $\om =0$ and also of width $\approx 1/(n\ta)$, is the background term, because being independent of $\ep_0$ the line is there even in the absence of the periodic signal.  However, when $\Om > 1/(n\ta)$ the signal line is located outside the Gaussian cutoff of this background line, and so the signal can still dominate the background even though the ratio of the former amplitude to the latter is $\approx \ep_0^4 \cT/\ta$ which might not be $\gg 1$.  This is consistent with our earlier claim that the second criterion of \eqref{crite} is also necessary for optimal signal detection.  Since the proposed algorithm of using $n\gg 1$ orthonormal filters enables the criterion to be satisfied more easily than the conventional single filter approach, it facilitates the recognition of embedded periodic or quasi-periodic transients.

\subsection{Noise in the power spectrum}

It remains to compare the fluctuation amplitude of the power spectrum to the strength of the resonance line.  As derived in Appendix B,
the variance of the power spectrum emerges as \bea {\rm var} (|\tilde\cI(\om)|^2) &=& \<\cI_b\>^4 \fr{\pi^2 \ep_0^4}{4}  \de^2 (\om-2\Om) \left[\fr{\sqrt{\pi}\ep_0^4 \cT\ta}{16} e^{-n^2 \ta^2 (\om-2\Om)^2/4} + \sqrt{\pi}\cT\ta  e^{-n^2 \ta^2 \om^2/4}\right]\notag\\
&& +2\<\cI_b\>^4 \left[\fr{\sqrt{\pi}\ep_0^4 \cT\ta}{16} e^{-n^2 \ta^2 (\om-2\Om)^2/4} + \sqrt{\pi}\cT\ta  e^{-n^2 \ta^2 \om^2/4}\right]^2. \label{8pt} \eea
In the absence of the periodic signal, \ie~when $\ep_0 =0$, the variance reduces to
\beq {\rm var} (|\tilde\cI(\om)|^2) = 2\pi \<\cI_b\>^4 \cT^2 \ta^2 e^{-n^2 \ta^2 \om^2/2}, \label{back} \eeq which is exponentially small at the frequency of the signal $\om=2\Om$ if $\Om > 1/(n\ta)$.

The ratio of the signal power to the standard deviation (standard error) in the power, or the first term of \eqref{lines} divided by the square root of the first term of \eqref{8pt}, is of order $\cT/\ta \gg 1$, and is independent of $n$.

\subsection{Simulation of the sensitivity enhancement}

A simulation is performed to test the performance of the orthonormal filters combination in detecting the embedded periodic signal of Section 4.2. The parameters are $\Om = 33.3$ kHz and $\ep_0^2 = 0.03$ for the signal, $\sigma_\om = 0.2$ MHz for the Gaussian spanned by the filter coefficients $\ba$, and a carrier wave frequency (central frequency of the Gaussian filter, or equivalently ${\om_0}$) of 10 GHz. The value of $\ep_0$ was chosen such that the signal is insignificant in the lower n-part series. For even smaller values of $\ep_0$ the overall result of the simulation does not change. Except that for successively smaller values of $\ep_0$ one would need to co-add more intensity series to find even weaker signals.

\begin{figure}[!htb]
\begin{center}
\includegraphics[width=6in]{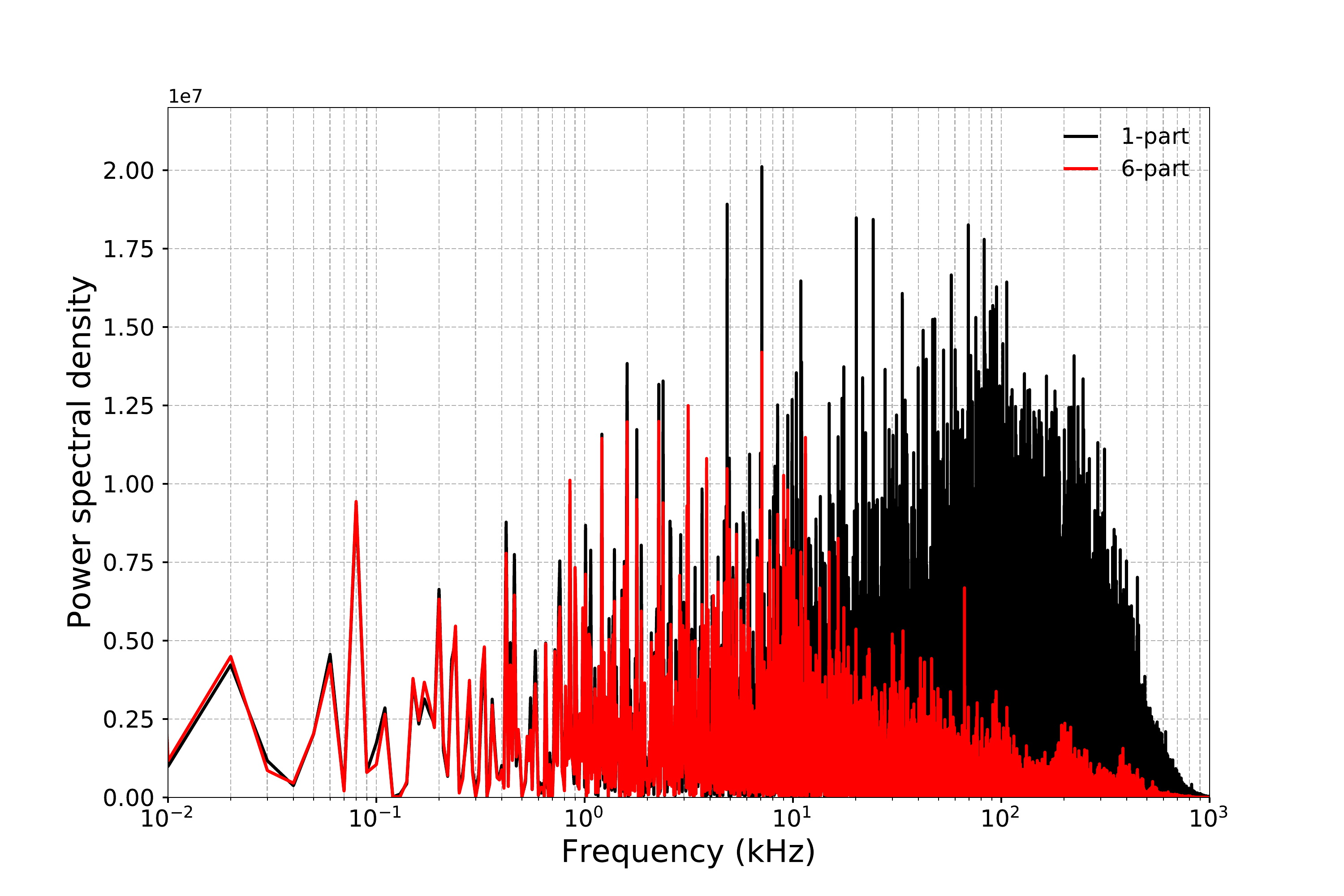}
\end{center}
\caption{The simulated power spectral density of 1-part (black) and overlapping 6-part (red). The signal power is prominent in 6-part at frequency $\omega$ = 66.6 kHz, \ie~at twice the frequency $\Omega$ = 33.3 kHz of the embedded sine wave. Table 2 contains 1-part to 6-part line significances measured from the power spectral density data and correlation lengths of the intensities.}
\label{sim_fig}
\end{figure}

Results of the simulation are summarized in Table 2 and Figure \ref{sim_fig}, where it can be seen in the former that the signal-to-noise of the (intensity) power spectral density excess at the frequency $2\Om$ (the finite frequency mode of $\sin^2 \Om t$) increases monotonically with the number of filters combined. The latter power spectrum shows the line is clearly identifiable within the noise after successive filtering.  Thus it can be seen that the simulation agrees well with the theory.

\begin{figure}[!htb]
\begin{center}
\includegraphics[width=6in]{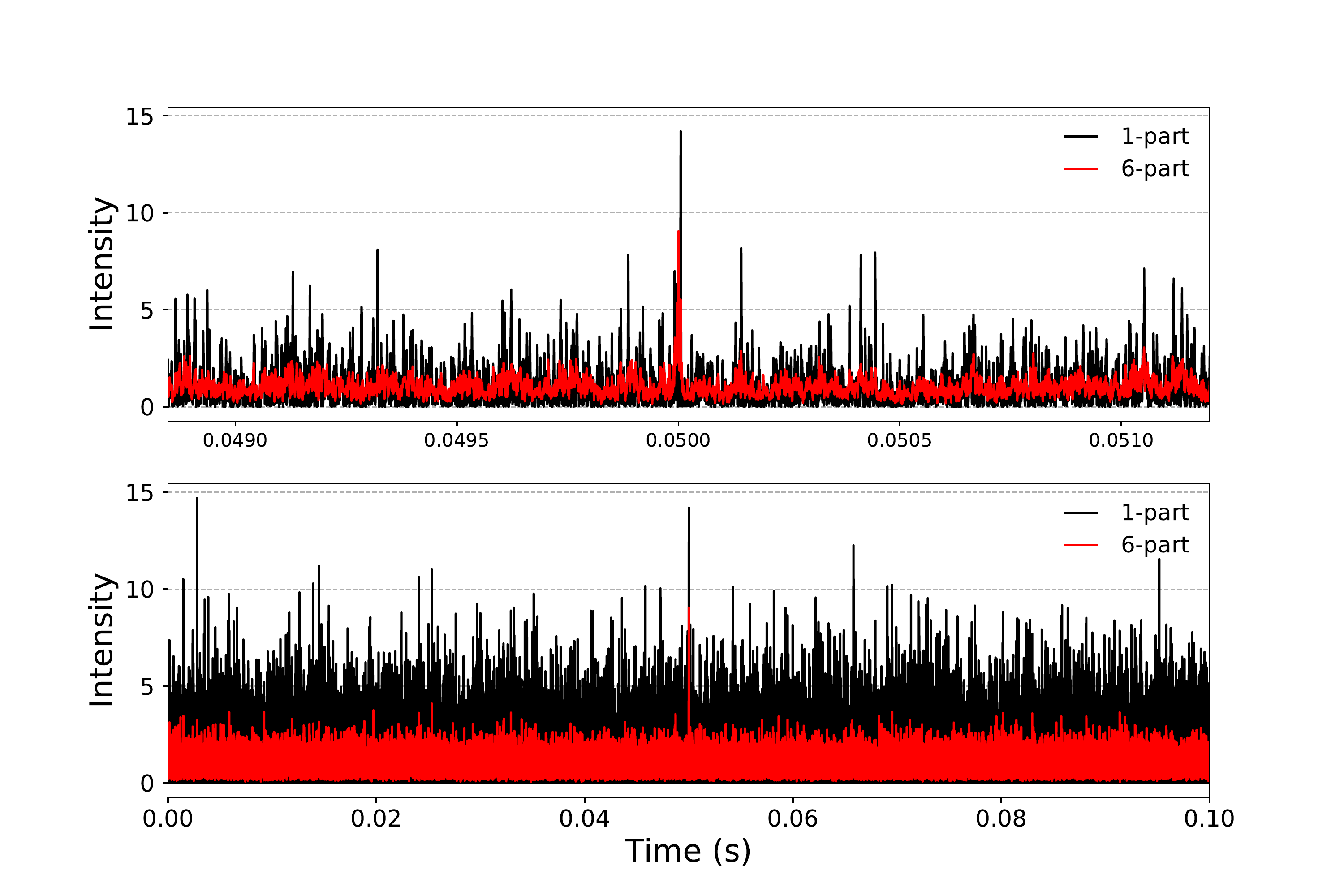}
\end{center}
\caption{The simulated intensity time series of 1-part (black) and overlapping 6-part (red). The transient Gaussian pulse is visible in the center of both figures. Increasing significance is shown in Table 3. The bottom figure shows the entire simulated time series, with Gaussian pulse embedded in the center. The signal-to-noise ratio of the Gaussian pulse increases roughly as $\sqrt n$. However, for even larger $n$ we find $\sqrt n$ sets a lower limit. For example, continuing to $n=21$, the SNR is measured to be $\sim$16 which is greater than the predicted signal-to-noise increase of $\sqrt{21}\times$(conventional $n=1$ SNR) $\sim 10$. Significances and SNRs for $n=1$ through $n=6$ are shown in Table 3.}
\label{sim_fig_pulse}
\end{figure}

The results of Section 4.2 apply to other time variable signals, for example a transient Gaussian pulse. The signal-to-noise ratio increases with the addition of $n$ orthogonally filtered intensity time series as $\sqrt{n}$. A Gaussian pulse of duration $\sim 5\tau$ embedded in Gaussian noise was simulated and results are shown in Figure \ref{sim_fig_pulse} and Table 3. 
\begin{table}[!htb]
\centering
\resizebox{12cm}{!} {
\begin{tabular}{|c|c|c|}
\hline
 Order of processing & Simulated line significance & Correlation length ($\mu s$)  \\
 & at $\om$=66.6 kHz = 2$\Om$ &   \\
\hline
1-part  & 2.8   &  12.0\\ \hline
2-part  & 1.8   &  24.0\\ \hline
3-part  & 1.7  & 36.0 \\ \hline
4-part & 2.9  & 48.0 \\ \hline
5-part & 7.4  & 60.0 \\ \hline
6-part & 8.1  & 72.0 \\ \hline
\hline
\end{tabular}}
\label{snr1}
\caption{Simulation results for a 33.3 kHz period signal embedded in Gaussian background noise. Intensity series for 1-part to 6-part were simulated. The correlation length of the intensity data, shown in last column, increases with filter order $n$ as $n\ta$, where $\ta$ is the coherence time of the 1-part series.}
\end{table}

\begin{table}[!htb]

%\caption{ }

\centering

\resizebox{12cm}{!} {

\begin{tabular}{|c|c|c|}

\hline

 Order of processing & Simulated pulse significance & Signal-to-noise ratio  \\

\hline

1-part  & 1.2   &  2.2\\ \hline
2-part  & 1.7   &  3.1\\ \hline
3-part  & 1.6  & 3.3 \\ \hline
4-part & 2.1  & 4.1 \\ \hline
5-part & 3.2  & 5.4 \\ \hline
6-part & 3.8  & 6.3 \\ \hline
    
\hline

\end{tabular}}

\label{signaltonoise_pulse}
\caption{Simulation results for a Gaussian pulse embedded in Gaussian background noise. Intensity series for 1-part to 6-part were simulated. Following the prediction of Section 4.2, the signal-to-noise ratio increases with filter order $n$ approximately as $\sqrt n$. The correlation lengths for 1-part to 6-part are the same as those found in Table 2.}
\end{table}

%\newpage
\section{Application to a VLA observation of PSR 1937+21}

We now apply the noise reduction algorithm to a 10 minute observation of the millisecond pulsar PSR 1937+21 by the VLA on 21:36 UTC October 29, 2015.  The voltage time series before processing comprise measurements at $T=15.625$~ns timing resolution.  In the Fourier domain, the modes cover a frequency range of 32 MHz with equal weights, centered at 1.4 GHz.  Owing to the wide frequency range, dispersion by the interstellar plasma is significant.  Yet, following the recipe of the previous section, we minimized dispersion effects by filtering the voltage modes digitally with a narrow Gaussian of 0.663 MHz FWHM, corresponding by \eqref{normcov} to a Gaussian intensity autocorrelation function of $400$~ns or $26$ sampling intervals FWHM, to produce the intensity time series $I_r$.  Specifically, from the published dispersion measure (DM) of $71$~pc~cm$^{-3}$ (\cite{kas94}, the dispersive broadening during propagation through the interstellar medium is of order \beq \de t = 142 \left(\fr{{\rm DM}}{71~{\rm pc}~{\rm cm}^{-3}}\right) \left(\fr{\De\nu}{0.663~{\rm MHz}}\right)
\left(\fr{1.4~{\rm GHz}}{\nu}\right)^3 \label{delay}~\mu{\rm s} \eeq for the relevant bandwidth, $0.663$~MHz, and is much smaller than the period of $1.56$~ms (\cite{kas94}).  Thus the pulsar light curve is unaffected within such a small $\De\nu$.

\subsection{Pulse profile}

\begin{figure}
\begin{center}
%\vspace{}
\includegraphics[width=6.5in]{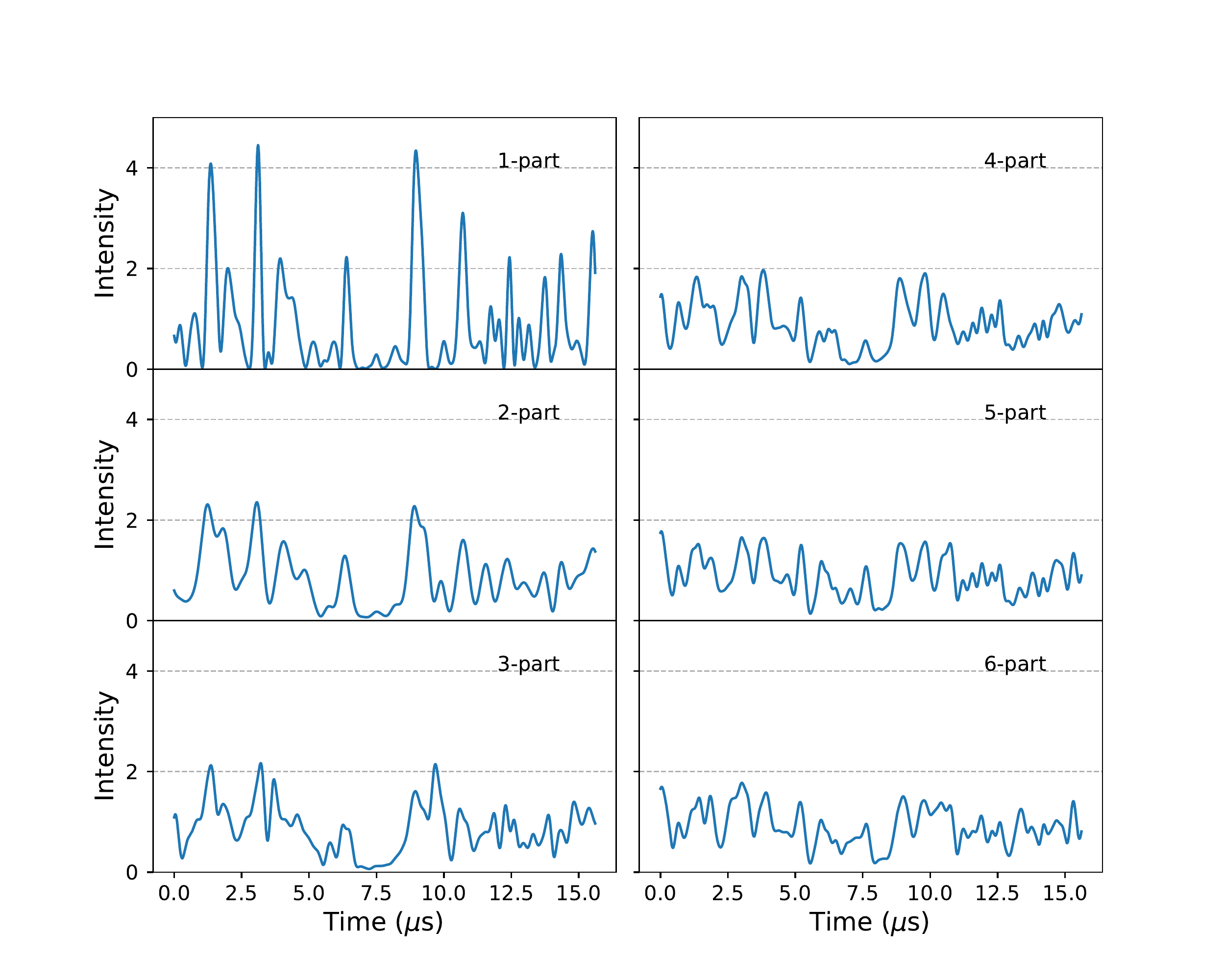}
\vspace{-1cm}
\end{center}
\caption{A small segment of the VLA intensity time series, with amplitude filtered digitally in the conventional way by a narrow Gaussian bandpass (1-part), and by $n$ multiple filters with $n\leq 6$.  The filters' designs were explained in Section 3 and illustrated in Figure \ref{filter6}.  The intensity of each series was normalized to a mean of unity, and the sampling time is $T=15.625$~ns.  Note the progressive decrease in the noise variance, accompanied by an increase in the noise coherence time, as $n$ increases.}

\label{one_to_six_intensity}
\end{figure}

When the voltage data were digitally processed by the $n$-part filtering algorithm outlined in Section 3 (with each part having the same Gaussian intensity bandwidth (see Figure \ref{filter6}) of 0.663 MHz), the resulting intensity time series are shown in Figure \ref{one_to_six_intensity} for $n= 1,2,\cdots ,6$, where it can be seen that the relative noise variance at the resolution $T$ comes down with increasing $n$.  Even more precisely, Table 1 shows that the relative variance is $1/n$, consistent with the theoretical prediction of Section 3, and the advantage of using an arbitrarily large $n$ goes away when the time series is averaged over indefinitely large time intervals $\cT =NT$ with $N \gg n$.  This too is consistent with theory, as is the broadening of the intensity autocorrelation function with $n$, Figure \ref{acf}. This provides the explanation of why the relative variance for $n>1$ goes down with decreasing $\cT$, Figure \ref{pulse_pro}.

To assess the performance in pulsar detection, the resulting $n$-part intensity time series were folded a modulo the $1.56$~ms period of PSR 1937+21.  In Figure \ref{pulse_pro} is shown a comparison of the pulse profiles of $n=1$ against $n=6$ at various timing resolution.  In accordance with expectation, it can be seen that the original (and highest) resolution data has the best signal-to-noise advantage in revealing the presence of pulsations when $n=6$.  The shape of the pulse profile is consistent with the primary and inter-pulse configuration reported in the literature (\eg~\cite{kas94}).  To quantitatively test whether the signal-to-noise enhancement of the pulse profile is real, we computed the ACF of the pulse profile for various $n$ ranging from $n=1$ to $n=5$, with ACFs for $n=1$ to $n=3$ shown in Figure \ref{acf}, and evaluated the statistical significance of the correlation using the Pearson $p$-statistic.  The results are presented in Tables 4 and 5 for raw (unbinned) data, as well as binned data to ensure all points are independent even for $n=5$.  It can be seen that for the unbinned data where the resolution is highest, the significance of the correlation is largest for $n=5$ and decreasing monotonically to $n=1$.  Same is true also for the binned data, although the improvement in the correlation coefficient is less because the noise reduction advantage of $n>1$ scenarios is meant to be absent (or substantially reduced) here.

Since our proposed methodology is best suited for detecting signals that vary on short timescales, but the pulsar modulation of the intensity occurs on timescales far larger than the ACF width of any of the $n$ orthogonal filter combinations employed, the significance enhancement with respect to$~n$ is indicative of the presence of periodic or quasi-periodic modulations on small timescales.

\begin{table}[ht]

%\caption{ }

\centering

\resizebox{12cm}{!} {

\begin{tabular}{|c|c|c|c|c|}

\hline

 Order of  & Autocorrelation  & Degrees of& Null hypothesis & Significance of \\
 processing & coefficient & freedom & acceptance probability & correlation \\

\hline

1-part & $0.135$  & $2392$ & $1.69 \times 10^{-11}$  & $6.83 \sigma$ \\ \hline
2-part  & $0.208$ & $1196$  &  $1.86 \times 10^{-13}$  & $7.45  \sigma$ \\ \hline

3-part  &  $0.256$  & $797$ & $1.08 \times 10^{-13}$ & $7.52  \sigma$ \\ \hline

4-part &  $0.310$ & $598$ & $4.36 \times 10^{-15}$ & $7.93  \sigma$ \\ \hline

5-part  & $0.349$ & $478$ & $1.94 \times 10^{-15}$ & $8.03 \sigma$ \\

\hline

\end{tabular}}

\label{signaltonoise}

\caption{The autocorrelation coefficient and statistical significance of the pulsar intensity profile in the original resolution of $15.625$~ns and at the time lag of 4,000 resolution elements, {\it viz.}~$62.5~\mu$s. Conversion from coefficient to significance was performed using the Student $t$-distribution and the Pearson $p$-statistic, see \eg~Chapter 14 of \cite{pre07}.  The degrees of freedom were evaluated by taking account of the larger correlation lengths of the $n$-part intensity series (Section 3 and Figure \ref{acf}).}

\end{table}

\begin{table}[ht]

%\caption{ }

\centering

\resizebox{12cm}{!} {

\begin{tabular}{|c|c|c|c|}

\hline

 Order of  & Autocorrelation  &  Null hypothesis & Significance of \\
 processing & coefficient &  acceptance probability & correlation \\

\hline

1-part & $0.497$  &  $9.38 \times 10^{-33}$  & $11.98 \sigma$ \\ \hline
2-part  & $0.512$  &  $6.75 \times 10^{-35}$  & $12.38  \sigma$ \\ \hline

3-part  &  $0.521$  & $3.09 \times 10^{-36}$ & $12.62  \sigma$ \\ \hline

4-part &  $0.539$ &  $3.64 \times 10^{-39}$ & $13.14  \sigma$ \\ \hline

5-part  & $0.579$ & $2.96 \times 10^{-46}$ & $14.33 \sigma$ \\

\hline

\end{tabular}}

\label{}

\caption{The autocorrelation coefficient and statistical significance of a binned pulsar intensity profile with the lower resolution of $3.125~\mu$s (or 200 original time bins) and at the time lag of $62.5~\mu$s. Conversion from coefficient to significance was performed using the Student $t$-distribution and the Pearson $p$-statistic, see the caption of the previous figure.  The degrees of freedom equal the number of data points that participated in the computation of correlation coefficient because the points are all independent of each other after binning.}
\end{table}
\subsection{Power spectrum of the pulse profile - microstructures}

Thus the next question is whether the suppression of noise in high timing resolution as offered by the proposed analysis technique could result in the discovery of fast transient emissions in the pulse profile that become averaged away in low resolution data.  To investigate, we computed the power spectral density (PSD) of the pulse profile, with ensuing frequencies necessarily quantized into multiples of the pulsar frequency.

\begin{figure}[!htb]
\begin{center}
%\vspace{}
\includegraphics[width=6in]{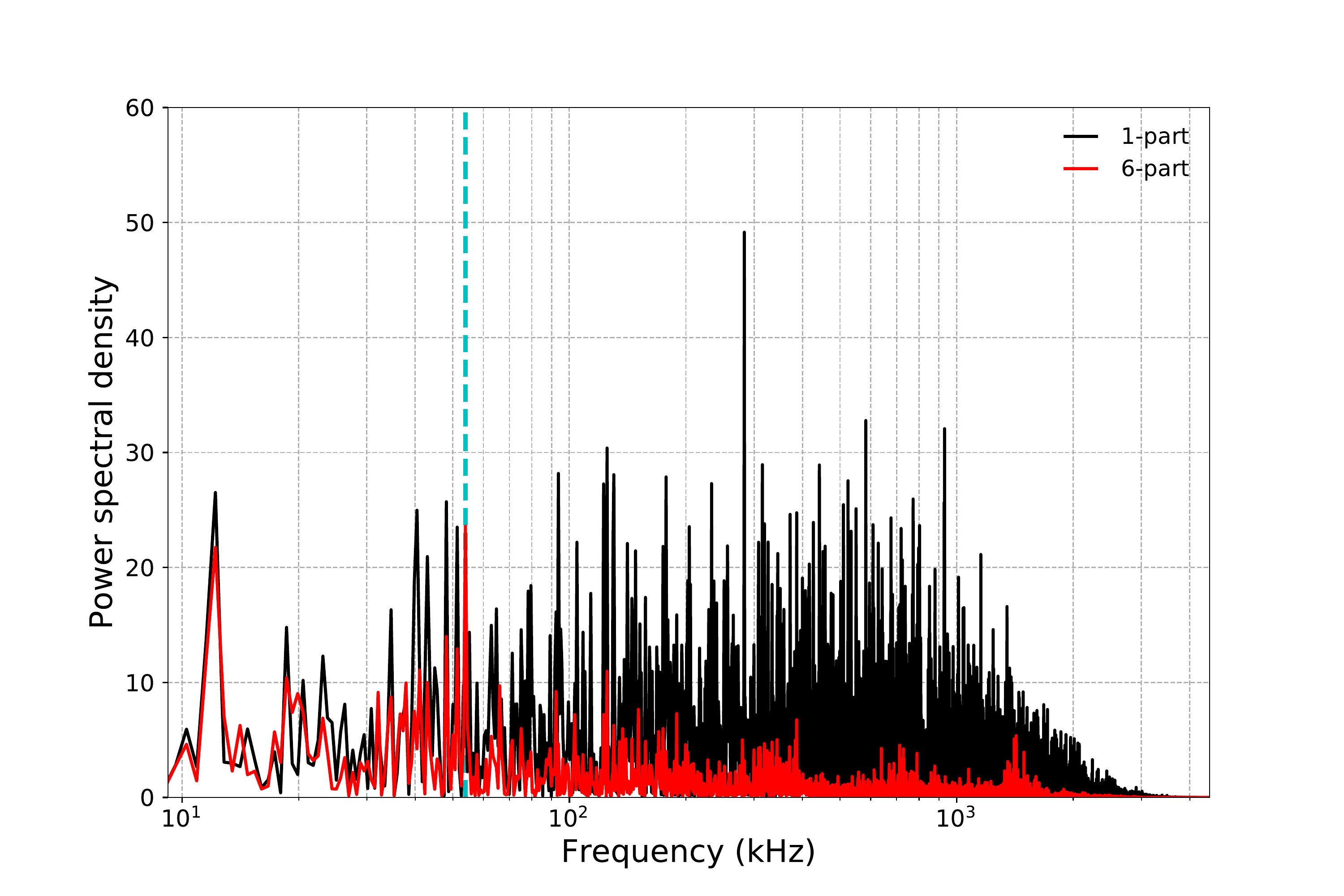}
%\vspace{-6cm}
\end{center}
\caption{The power spectral density of ten minutes of stacked pulsar data, 1-part (black) and overlapping 6-part (red). The vertical dashed line shows the location of the 53.9 kHz pulsar harmonic.}

\label{psd_har}
\end{figure}

In Sections 4.1 $-$ 4.4 and the simulation of Section 4.5, we showed the variance of the noise spectral density decreases with filter order $n$. In Figure \ref{psd_har} we observe this behavior in the power spectra of the filtered and stacked intensity data. The spectrum frequency spacing is $\om_p$, where $\om_p$ is the pulse frequency. Correspondingly, every peak in the spectrum is the location of a potential pulsar harmonic line.

By \eqref{delay} the time delay across the entire bandwidth 
$\De\nu$ of 0.663 MHz is $\approx 142~\mu$s. 
Slow modulations that take place on timescales $\gg 1/\De\nu$ will survive dispersion.
While it is the case that modulations occuring on timescales nearing $1/\De\nu$, 
where $\De\nu \approx 0.7$~MHz, are affected by dispersion of the interstellar 
medium, this is applicable to modulaitons comprised of frequency components spanning $\De\nu$. 
 
\begin{figure}[!htb]
\begin{center}
%\vspace{}
\includegraphics[width=5.5in]{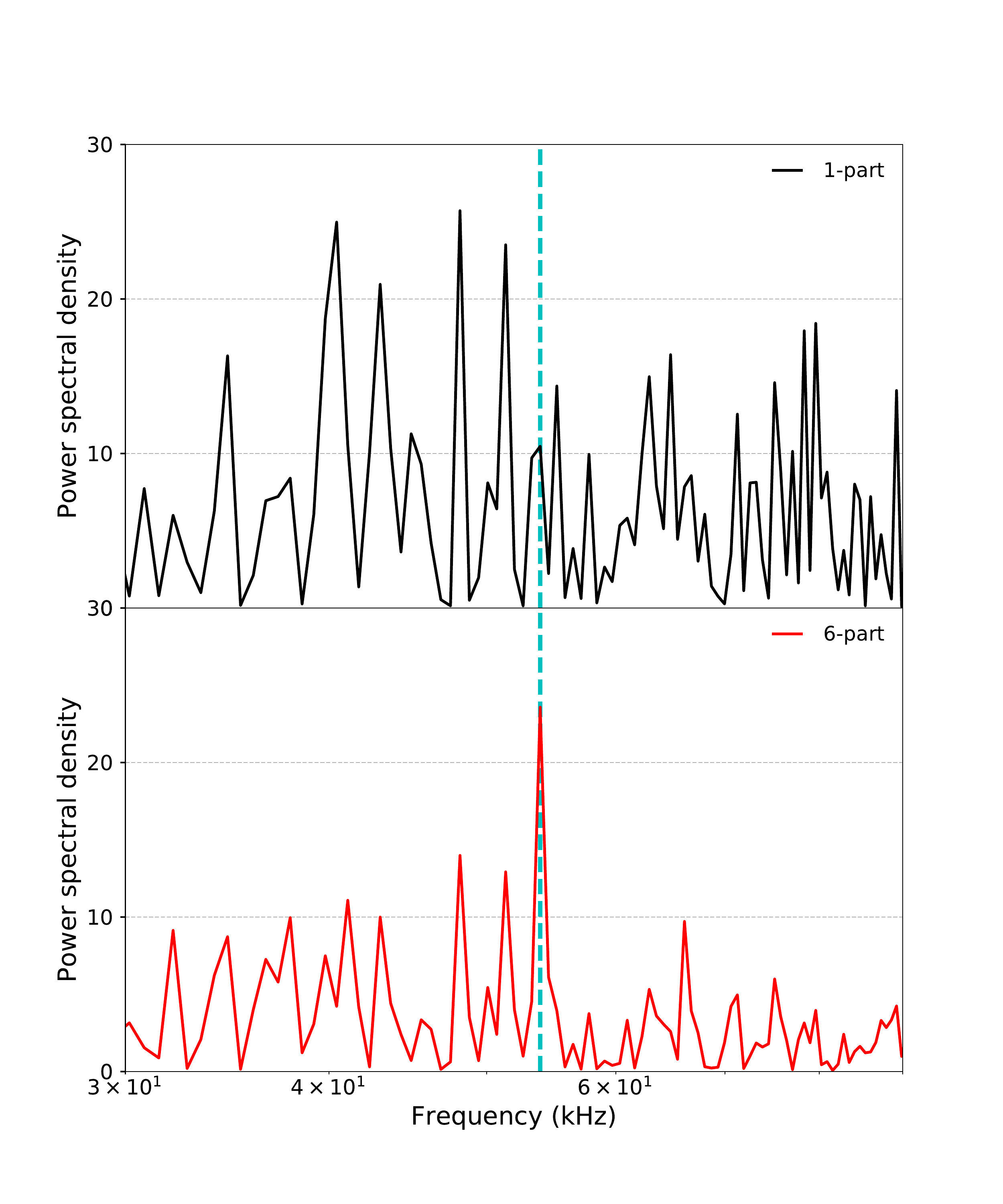}
%\vspace{-6cm}
\end{center}
\caption{The top and bottom figures are, respectively, the 1-part and 6-part power spectral density of ten minutes of pulse profile data. The line is insignificant in 1-part and one must process the data with digital filters to `uncover' it. Table 6 contains line significances for 1-part through 6-part. The 53.918 kHz harmonic line is highlighted with a vertical dashed line.}

\label{har0}
\end{figure}

%The frequency spread of faster modulations with frequency component spanning a much smaller fraction 
%of the observer's $\De\nu$, are able to survive dispersive broadening. 
Faster modulations with a frequency spread that is a sufficiently smaller fraction 
of the observer's $\De\nu$ will survive dispersive broadening. 
By visual inspection of the pulse profile PSD we identified one such candidate modulation, 
the 83$^{\rm rd}$ pulsar harmonic line. We subsequently tested the line against the theory. 
Table 6 contains the line significance as a function of filter order, 
from the PSD of ten minutes of stacked profile data, up to 6$^{\rm th}$ processing order. 
The line is insignificant in 1-part, but the noise is sufficiently reduced to reveal an 
$\sim 8\sigma$ line in 6-part. Figure \ref{har0} shows that the harmonic line is buried in the noise 
in 1-part and becomes prominent as the noise is suppressed, in 6-part. 

On the other hand, the lowest frequency region of the pulse profile PSD is dominated in 
power by the lower-order pulsar harmonic lines. There is not enough background data in this 
regions to properly test these lines against the theory.  This prevents the search of lower frequency harmonic lines.

\begin{table}[!htb]

%\caption{ }

\centering

\resizebox{12cm}{!} {

\begin{tabular}{|c|c|c|}
 \hline
 Order of processing & 53.9 kHz line significance & Correlation length ($\mu s$)  \\

\hline

1-part  & 0.55  &  0.6250  \\ \hline
2-part  & 0.73  &  1.250  \\ \hline
3-part  & 1.64 & 1.875 \\ \hline
4-part & 2.66  & 2.500  \\ \hline
5-part & 4.96  & 3.125  \\ \hline
6-part  & 7.84  &  3.750  \\ \hline

\hline

\end{tabular}}

\label{harmonic1}

\caption{Significance of the 83$^{\rm rd}$ pulsar harmonic line, as a function of filter order, from the power spectral density of ten minutes of stacked intensity data. Spectra for 1-part and 6-part are compared for ten minutes of profile data in Figure \ref{har0}.}

\end{table}

\newpage
\section{Application to LIGO Binary Black Hole Merger Event GW150914}
Another application of the `eigen-filter' (orthogonal filtering) algorithm presented above attempts to increase the search sensitivity of transient gravitational-wave events. Short-lived bursts of GWs are the primary target application; in particular, bursts of GW radiation detectable over a few coherence times, which are unanticipated, \ie~where the theoretical waveform is unknown.

 In the first and second observing runs, the Laser Interferometer Gravitational-Wave Observatory (LIGO; \cite{ab92}) in the US and Italy's Virgo (\cite{ac12}) have confidently detected transient GWs from sources with known waveform models, namely merging black hole binaries (BBH) and merging neutron stars (\cite{lv19}). Japan's Kamioka Gravitational-Wave Detector (KAGRA; \cite{as13}) is scheduled to join in the latter part of the third observing run, O3, which concludes April 30, 2020. At times and frequencies where non-stationary, non-Gaussian noise sources are absent, for example glitches, the LIGO data streams are statistically locally stationary and Gaussian \cite{ab19}. Noise of this character masks weak embedded GW signals, of any form, at all times, but it is precisely this time-independence that permits its systematic reduction. 

One method to find weakly modeled, or unmodeled candidate GW bursts, is with analysis algorithm Coherent WaveBurst (cWB) (\cite{ki08} and \cite{ki16}), currently in use by LIGO-Virgo collaboration. The cWB algorithm incorporates both excess power and cross correlation between detector pairs to identify triggers. Here we show excess power stands to gain a $\sim \sqrt{n}$ increase over Gaussian noise power.

This is a new  facility for coherent (multi-detector) searches.  It enlists eigenfiltered intensity time series, 
where we show that cross correlation significance of detector intensities increases with filter order $n$ 
for a real signal. We argue below that increased significance serves to decrease the false alarm rate.  
It is hoped that the technique presented here would improve the sensitivity and efficacy of low latency pipelines.

Of course, the technique presented here in no way renders any less critical the work performed to exclude as astrophysical non-stationary, non-Gaussian transient noise events. For ground-based detectors, these noise sources can be the cause of global-scale environmental influences and detector noise. For example, coincident noise triggers, which are not infrequent, may also correlate in time. (\cite{ab16}) 

As proof-of-principle, we apply eigen-filtering to the first BBH merger event, GW150914, detected by LIGO (\cite{ab16a}). We use 32 seconds of publicly available LIGO strain data surrounding the event from Livingston and Hanford, obtained from the \cite{gwosc}. The data have bandwidth 2048 Hz and are sampled at the Nyquist rate, 4096 Hz. GW150914 falls under the category of transients whose SNR is maximized by applying a matched filter, constructed from known waveform templates, to the data streams of each detector separately. For that reason, SNRs for GW150914 reported in the literature are larger than those computed from excess power alone. 

\begin{figure}
\begin{center}
%\vspace{}
\includegraphics[width=6.5in]{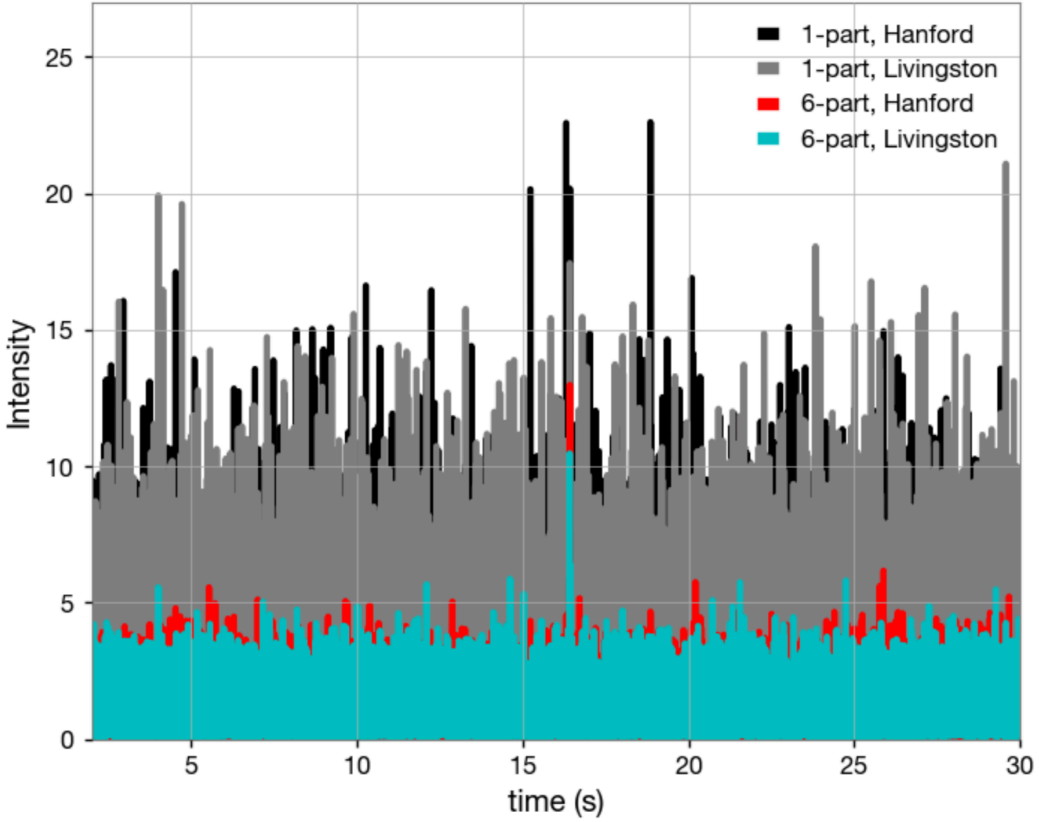}
%\vspace{-6cm}
\end{center}
\caption{Intensity time series of 1-part (Hanford, black; Livingston, grey) and overlapping 6-part (Hanford, red; Livingston, cyan). The transient GW event clearly rises above the noise in 6-part.The signal-to-noise ratio of the event increases roughly as $\sqrt n$}

\label{gw150914_inten}
\end{figure}

First, we describe the data pre-conditioning and how the eigen-filters are constructed for specific application to LIGO data, the strain amplitude data are first whitened, so $\si_I^2/\bar{I}^2$, for the $n^{\rm th}$ intensity series, is $\sim 2/n$.  Here, strain amplitude data are analogous to radio voltage amplitudes in the preceeding sections, $V(t)$. Symmetry of the eigen-filters across the BW of interest is required to satisfy condition \eqref{zero} as close to identically as possible. A `first order approach' to enforcing symmetry, however, is to apply an eigen-filter with a boxcar profile to the original BW, \ie~we do not bandpass the data. For our demonstration, this is an acceptable starting point. One must also keep in mind, any applied filter BW must be $\gg$ than the signal frequency for eigen-filtering to successfully corroborate theoretical prediction. For GW150914 the signal frequency spans approximately one decade, between 30 and 300 Hz. After these two operations, we measure a finite non-zero correlation length that increases with filter order $n$ as $n\tau$, where $\tau$ is the correlation length of the $n$=1 intensity series. Results for the Hanford and Livingston event significance and SNR as a function of filter order are shown in Table 7 and Figure \ref{gw150914_inten}. 

\begin{figure}
\begin{center}
%\vspace{}
\includegraphics[width=6.5in]{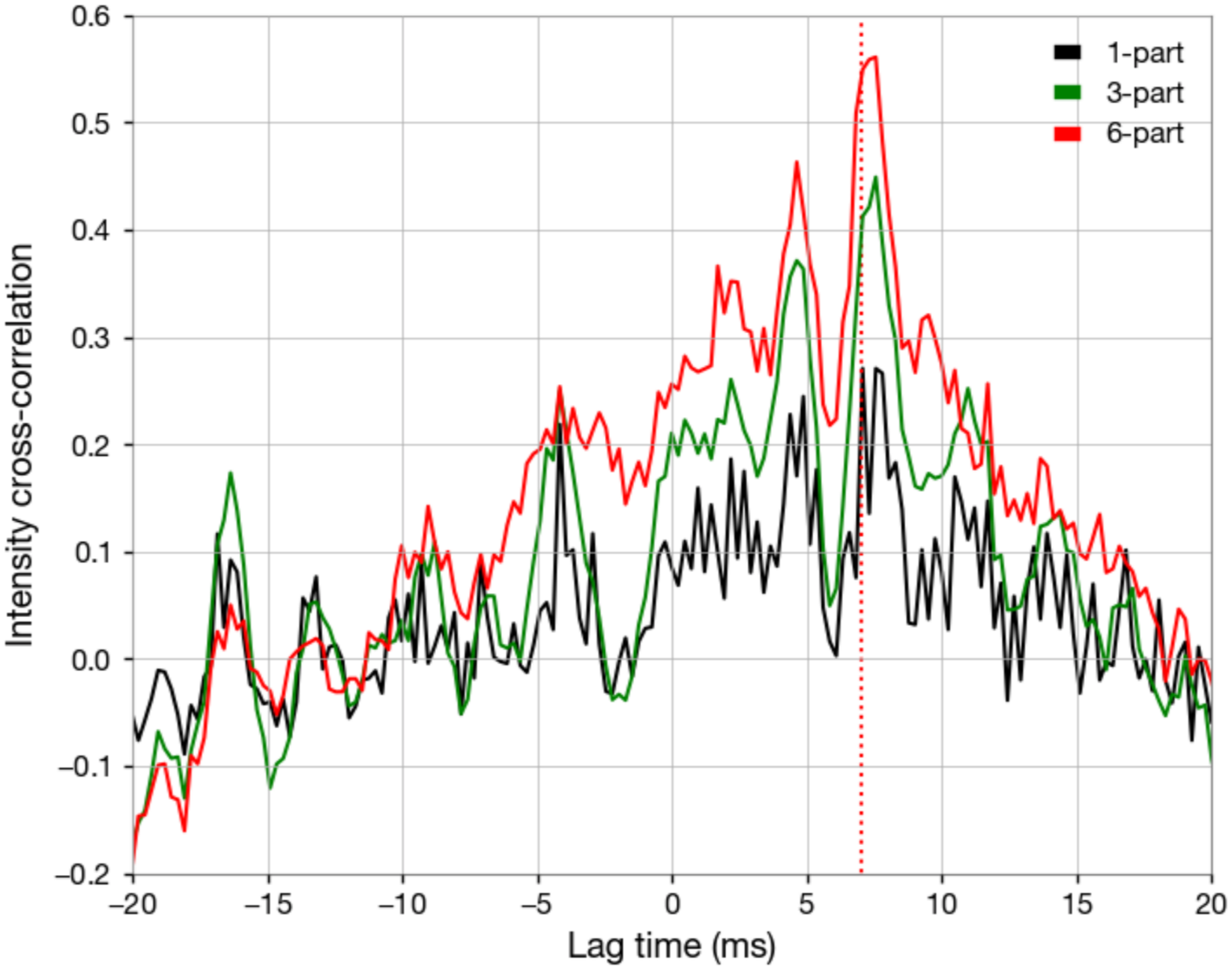}
%\vspace{-6cm}
\end{center}
\caption{The 0.1 seconds leading up to the end of merger were cross correlated between Hanford and Livingston 
eigen-filtered intensity time series. Correlation increases with filter order $n$. 
Respectively, black, green and red curves correspond to co-added intensity series 
$n$=1,3 and 6. The vertical dashed red line marks the 7 ms lag time, which is within the region of peak cross correlation 
for each intensity series.} 

\label{gw150914_cross}
\end{figure}

We also cross correlate LIGO Hanford and Livingston eigen-filtered intensity data streams for a total of 0.1 seconds preceding the reported `merger end time'. The known delay time between Hanford and Livingston is recovered. The merger event in Hanford's detector stream lags $\sim$7 ms behind that in Livingston's. Increased correlation with filter order is shown in Figure \ref{gw150914_cross}. The same result holds when larger segments of data are cross correlated.

\begin{table}[ht]

%\caption{ }

\centering

\resizebox{16cm}{!} {

\begin{tabular}{|c|c|c|c|c|c|}

\hline

 Order of processing & Significance  & Signal-to-noise ratio  & Significance & Signal-to-noise ratio  & Correlation length\\
 & Hanford  &Hanford  & Livingston & Livingston &  (ms) \\

\hline

1-part  & 1.4   &  2.1 & 0.9 & 1.6 &  0.2 \\ \hline
2-part  & 1.9   &  2.9 & 1.3 & 2.3   & 0.4  \\ \hline
3-part  & 2.3  & 3.6 & 1.6 & 2.8  & 0.6   \\ \hline
4-part & 2.7  & 4.1 & 1.8 & 3.2 & 0.8  \\ \hline
5-part & 2.9  & 4.5 & 2.0 & 3.5 & 1.0   \\ \hline
6-part & 3.1  & 4.9 & 2.1 & 3.9 &  1.2 \\ \hline
    
\hline

\end{tabular}}

\label{gw150914pulse}
\caption{GW150914 excess power signal-to-noise ratios and significances, for co-added intensity series processed by $n=1$ to $n=6$ eigen-filters. SNR increases approximately as $\sqrt{n}$. Correlation length increases as $n\tau$, where $\tau$ is the correlation length of the $n$=1 intensity series. The SNR is computed as the mean intensity over noise for 30 ms surrounding the peak intensity, specifically between UTC 1126259462.41 and 1126259462.44 for Hanford. The same length interval is used for Livingston, but with averaging window shifted 7 ms to earlier times, corresponding to the GW's earlier arrival time at the Livingston detector.}
\end{table}

The question of how the false alarm rate (FAR) is affected by eigenfiltering is 
addressed next. Suppose a weak GW event, similar to GW151226, lasts one second and a threshold of 3 
sigma is set to find it.  This means in every 1,000 seconds of data one expects to receive 1 false alarm 
(3.3 sigma is approximately 1 in 1,000). Thus, if there are two time series each lasting 1,000 seconds, 
1 false alarm event in each is expected.  The probability of the two events to also be time coincident 
to within 1 ms, say,  is 1 ms/1000 s $\approx 10^{-6}$ .  
 With the CCF criterion included, we expect $10^{-6}$ burst in 1000 s, or a FAR of 
 0.03/yr. This estimate is on par with that given by coherent WaveBurst for GW151226 (\cite{lv19}).  
 With the use of eigenfilters, the significance for an actual GW event increases with filter order. 
 This will lead to a smaller FAR than just estimated, setting an upper limit with no eigenfiltering.

\section{Conclusion}

An algorithm is proposed to significantly reduce the Gaussian noise of radio intensity time series by 
digitally designing $n$ amplitude filters, where $n\geq 1$ is arbitrarily large, having identical intensity 
bandwidths such that the resulting co-added intensity time series has relative variance $\approx 1/n$ 
on timescales $\lesssim$ the coherence time of the noise, but recovers to the limit set by the radiometer 
equation \eqref{radio} in the opposite limit of long timescales as required by the 
Cramers-Rao bound (\cite{nai15}).

Therefore the primary caveat is the 
sensitivity limit imposed by the radiometer equation \eqref{radio}.  
The signal-to-noise ratio gradually drops back to the conventional value given by \eqref{radio}, 
as the integration time of the intensity increases. 
As one increases the number of eigenfilters, a reduction in the 
relative variance of the intensity will be maintained for longer integration times.
The limit of \eqref{radio} cannot be surpassed, so one should 
not use eigenfiltering to enhance variations existing on timescales much greater than the coherence time 
of the radiation. In this case, averaging the data gives the greatest advantage,
where the relative variance of the intensity is constrained by \eqref{radio}.

The method is applied to a 10 minute VLA observation of the millisecond pulsar PSR 1937+21 at the resolution of $T=15.625$~ns and FWHM intensity bandwidth $\De\nu = 0.663$~MHz.  It is found that even when $n$ is as low as $n\approx 5$ the pulsar intensity profile has much lower noise than the conventional single filter scenario (for equal $\De\nu$ in both cases) unless the time series are bin averaged to a resolution much poorer than $n/\Delta \nu \approx n\tau$. In this way, faint transient signals that occur on timescales between $\tau$ and $n\tau$ are enhanced with respect to noise.
The existence of a 53.9 kHz periodic modulation (corresponding to the 83$^{\rm rd}$ pulsar harmonic) in the intensity pulse profile is revealed only by applying n$>$3 filter combinations.

Since the advantage exists only at high resolution, the algorithm is best suited to the search of faint and fast transients that would otherwise be smeared out by any noise suppression scheme involving time averaging.  

Lastly we applied the algorithm to the first gravitational wave event detected by LIGO. We found the intensity signal-to-noise ratio of the event increases roughly as $\sqrt n$. We demonstrate the cross correlation between the Hanford and Livingston intensity series for 0.1 s preceding the merger's end time increases with filter order $n$.

While our focus has been on astrophysical signals, in principle, 
eigenfiltering is applicable to any type of Gaussian noise 
limited signal, including artificial (or non-astrophysical) signals. In future work, we will apply the algorithm 
to the most distant man-made signal received at Earth, from the Voyager I spacecraft. 
On and off target Voyager I baseband data, obtained with the 
Robert C. Byrd Greenbank Telescope in West Virginia, are accessible through the
Breakthrough Listen project public archive\footnote{https://breakthroughinitiatives.org/opendatasearch} 
\cite{wo18}. 

%There is another potential application of this algorithm, involving gravitational-wave (GW) sources detectable by the future space-based Laser Interferometer Space Antenna (LISA; \cite{am17}). A significant source of astrophysical noise to LISA at mHz frequencies is a confusion noise from tens of millions of unresolved ultra-compact Galactic binaries (UCBs) (\cite{lit19}). The superposition of many UCBs, each emitting primarily monochromatic GWs at a particular frequency and random phase, constitute an effective stochastic foreground noise, that must be accounted for when attempting to sift and resolve extragalactic signals occupying the same frequency band. The superposed UCB signals constitute a Gaussian noise time series with faint, resolvable GW signals embedded. Moreover, Figure 1a of \cite{lv16} implies an extragalactic stochastic GW binary black hole background will be present in the LISA data stream, \ie~if one extends the spectrum into the LISA band as $f^{2/3}$. In theory, these foreground and background Gaussian noise sources are reducible in GW intensity time series with the algorithm presented above.

The authors are grateful to Paul Demorest, Barry Clark, and Jean Eilek at NRAO Socorro for helpful discussions, and to Paul Demorest for providing the VLA data of PSR 1937+21. KL's research was supported by an appointment to the NASA Postdoctoral Program at the NASA Marshall Space Flight Center, administered by Universities Space Research Association under contract with NASA.
%KL is also grateful to Peter Bender for bringing attention to a population of unresolved extragalactic binary black holes as a potential source of noise for LISA.

%
\appendix
\section{Validity of the radiometer equation as applied to long term intensity averaging}

When the narrow band filter coefficients $\{a_j\}$ digitally multiply the voltages of the incident radiation with a flat intensity spectrum, and the exercise is repeated using another set of narrow band coefficients $\{b_j\}$, the two ensuing intensity series were denoted by $I_r$ and $I'_r$ in Section 3, while the summed intensity $\cI_r = I_r + I'_r$ was shown to possess 50 \% less relative variance then $I_r$ or $I'_r$ individually.  Despite this apparent advantage the radiometer equation governing the relative variance of the sample mean intensity taken over many contiguous coherence times was shown to remain valid under the scenario of two specific Gaussian-type filters $\{a_j\}$ and $\{b_j\}$.  In this Appendix we demonstrate the validity of the radiometer equation for {\it any} filters $\{a_j\}$ and $\{b_j\}$.

We begin with by substituting \eqref{covhalf} into \eqref{vc} and evaluating one of the double summations as an integral (the other summation then assumes the value $N$), {\it viz.}~\bea \var(\cI_\cT) &=& \fr{1}{N^2}\sum_{r,s=1}^{N} \cov(\cI_r,\cI_s) \notag\\
&=& \fr{\cT^2}{NT} \int \{[a(\om) a^*(\om')]^2 + [b(\om) b^* (\om')]^2 + 2a(\om)a^* (\om') b(\om)b^* (\om')\}e^{-i(\om-\om')t} d\om d\om' dt \notag\\
&=& \fr{2\pi\cT^2}{NT} \int \{[a(\om) a^*(\om')]^2 + [b(\om) b^* (\om')]^2 + 2a(\om)a^* (\om') b(\om)b^* (\om')\} \de (\om-\om') d\om d\om' \notag\\
&=& \fr{2\pi\cT^2}{NT} \int [|a(\om)|^2 + |b(\om)|^2]^2  d\om \notag\\
&\approx& \fr{\ta}{NT} (\<I_r\> + \<I'_r\>)^2,
\label{predelta} \eea
In arriving at the last step use was made of \eqref{meanI} and the approximation  \beq \cT^2 \int |a(\om)|^4 d\om \approx \fr{\<I_r\>^2}{\De\om} \approx \<I_r\>^2 \ta. \eeq  Thus $\var (\cI_\cT)/\<\cI\>^2 \approx \ta/(NT)$, consistent with the radiometer equation \eqref{radio}.

\section{Power spectrum of the intensity and its noise}

The power spectral amplitude $\tilde\cI(\om)$ of intensity fluctuation is the Fourier transform of the (stochastic) intensity time series $\cI(t)$.  The modulus square of $\tilde\cI(\om)$ is the power spectrum \beq |\tilde\cI(\om)|^2 = \int dt_2 \int dt_1 ~e^{i\om (t_2 - t_1)} \cI(t_1) \cI(t_2). \label{ap} \eeq  Changing the variable $t_2$ to $t$ where $t = t_2 - t_1$, and taking the expectation (ensemble average) value, we have
\bea \<|\tilde \cI (\om)|^2 \> &=&  \int dt_1~e^{-i\om t_1} \int dt_2~e^{i\om t_2} [\<\cI(t_1) \cI(t_2)\> - \<\cI (t_1) \> \<\cI (t_2) \> ] \notag\\
&& + \int dt_1~e^{-i\om t_1} \int dt_2~e^{i\om t_2} \<\cI (t_1) \> \<\cI (t_2) \> \notag\\
\label{apowspec}
\eea
where, by \eqref{Iex} and such equations as \eqref{varex} and \eqref{covex},
\beq \<\cI(t_1) \cI(t_2)\> - \<\cI (t_1) \> \<\cI (t_2) \> = \xi (t_2 - t_1) (1+\ep_0^4 \sin^2 \Om t_1 \sin^2 \Om t_2), \label{aI2pt} \eeq
and \beq \<\cI (t_1) \> \<\cI (t_2) \> = \<\cI_b\>^2 (1+\ep_0^2\sin^2 \Om t_1 +\ep_0^2\sin^2 \Om t_2 + \ep_0^4 \sin^2 \Om t_1 \sin^2 \Om t_2), \label{It1It2} \eeq
with $\xi (t)$ as defined in \eqref{xi}.

The subsequent evaluation of the integrals was already performed in that portion of the main text around \eqref{xi}, resulting in \eqref{lines} as the expression for $|\tilde\cI(\om)|^2$ in the vicinity of $\om = 2\Om$.  Specifically at and around $\om=2\Om$, \bea \int dt_1~e^{-i\om t_1} \int dt_2~e^{i\om t_2} {\rm cov}~[\cI (t_1),\cI(t_2)] &=& \<\cI_b\>^2 \left[\fr{\sqrt{\pi}\ep_0^4 \cT\ta}{16} e^{-n^2 \ta^2 (\om-2\Om)^2/4} + \sqrt{\pi}\cT\ta  e^{-n^2 \ta^2 \om^2/4}\right]\notag\\
&& + \cdots~, \label{2FT2pt} \eea and \beq \int dt_1~e^{-i\om t_1} \int dt_2~e^{i\om t_2} \<\cI (t_1) \> \<\cI (t_2) \> = \<\cI_b\>^2 \fr{\pi^2 \ep_0^4}{16}  \de^2 (\om-2\Om) +~\cdots~, \label{2FTII} \eeq
where ${\rm cov}~[\cI (t_1),\cI(t_2)] = \<\cI(t_1) \cI(t_2)\> - \<\cI (t_1) \> \<\cI (t_2) \>$ the missing terms represented by `$\cdots$' apply only to the scenario of $\om =0$, a frequency far away from $2\Om$, via at least one factor of $\de (\om)$ in these terms.

To calculate the expectation value of the variance of $|\tilde \cI (\om)|^2$, {\it viz.}~ \bea {\rm var} (|\tilde\cI(\om)|^2) &=& \<\tilde \cI (\om) \tilde \cI^* (\om) \tilde \cI (\om) \tilde \cI^* (\om) \> - \<\tilde \cI (\om) \tilde \cI^* (\om)\>^2, \notag\\
&=& \int dt_1~e^{-i\om t_1} \int dt_2~e^{i\om t_2} \int dt_3~e^{-i\om t_3} \int dt_4~e^{-i\om t_4} \<\cI (t_1) \cI (t_2) \cI (t_3) \cI (t_4)\> \notag\\
&& -~\<\tilde \cI (\om) \tilde \cI^* (\om)\>^2, \label{4ptI}
\eea
where $\<\cI (t_1) \cI (t_2) \cI (t_3) \cI (t_4)\>$ is a voltage 8-point function of the form $\<V_1 V_2 V_3 V_4 V_5 V_6 V_7 V_8\>$ with $V_1 = V(t_1)$, $V_2 = V^* (t_1)$, $V_3 = V(t_2)$, $V_4 = V^* (t_2)$~{\it etc}.  In this notation only 10 of the 24 different contraction patterns contribute significantly to the integral of \eqref{4ptI} at frequencies in the vicinity of $\om = 2\Om$ with $\Om$ satisfying \eqref{crite}.  They are as follows.

Firstly, one single contraction $(1,2)(3,4)(5,6)(7,8)$ that yields $\<\cI (t_1)\> \<\cI (t_2)\>\<\cI (t_3)\>\<\cI (t_4)\>$, contributing the quantity \beq \alpha = \<\cI_b\>^4 \left[\fr{\pi^2 \ep_0^4}{16} \de^2 (\om - 2\Om)\right]^2 \label{alpha} \eeq  to the integral of \eqref{4ptI}.

Secondly, six contractions $(1,2)(3,4)(5,8)(6,7)$, $(1,4)(2,3)(5,6)(7,8)$, $(1,2)(3,6)(4,5)(7,8)$, $(1,6)(2,5)(3,4)(7,8)$, $(1,8)(2,7)(3,4)(5,6)$, and $(1,2)(3,8)(4,7)(5,6)$, each yielding a term of the form $\<\cI (t_1)\> \<\cI (t_2)\> {\rm cov}~[\cI (t_3), \cI (t_4)]$ or $\<\cI (t_3)\> \<\cI (t_4)\> {\rm cov}~[\cI (t_1), \cI (t_2)]$ or other arrangements, and anyone of such terms contribute  the quantity \beq \beta = \<\cI_b\>^4 \fr{\pi^2 \ep_0^4}{16}  \de^2 (\om-2\Om) \left[\fr{\sqrt{\pi}\ep_0^4 \cT\ta}{16} e^{-n^2 \ta^2 (\om-2\Om)^2/4} + \sqrt{\pi}\cT\ta  e^{-n^2 \ta^2 \om^2/4}\right] \label{beta} \eeq  to the integral of \eqref{4ptI}.

Thirdly, three contractions $(1,4)(2,3)(5,8)(6,7)$, $(1,8)(2,7)(3,6)(4,5)$, and  $(1,6)(2,5)(3,8)(4,7)$, yielding ${\rm cov}~[\cI_1,\cI_2)] {\rm cov}~[\cI_3,\cI_4)]$ and two other arrangements, and each of these three terms contribute the quantity \beq \gamma = \<\cI_b\>^4 \left[\fr{\sqrt{\pi}\ep_0^4 \cT\ta}{16} e^{-n^2 \ta^2 (\om-2\Om)^2/4} + \sqrt{\pi}\cT\ta  e^{-n^2 \ta^2 \om^2/4}\right]^2 \label{gamma} \eeq to the integral of \eqref{4ptI}.

Finally, when the last term of \eqref{4ptI}, whose value is given by squaring the right side of \eqref{lines}, is subtracted from the quantity $\alpha + 6\beta + 3\gamma$ which is the sum of all the contributions the various contractions of $\<V_1 V_2 V_3 V_4 V_5 V_6 V_7 V_8\>$ made to the integral of\eqref{4ptI}, the variance of the power spectrum emerges as \bea {\rm var} (|\tilde\cI(\om)|^2) &=& \<\cI_b\>^4 \fr{\pi^2 \ep_0^4}{4}  \de^2 (\om-2\Om) \left[\fr{\sqrt{\pi}\ep_0^4 \cT\ta}{16} e^{-n^2 \ta^2 (\om-2\Om)^2/4} + \sqrt{\pi}\cT\ta  e^{-n^2 \ta^2 \om^2/4}\right]\notag\\
&& +2\<\cI_b\>^4 \left[\fr{\sqrt{\pi}\ep_0^4 \cT\ta}{16} e^{-n^2 \ta^2 (\om-2\Om)^2/4} + \sqrt{\pi}\cT\ta  e^{-n^2 \ta^2 \om^2/4}\right]^2. \label{8pt2} \eea
In the absence of the periodic signal, \ie~when $\ep_0 =0$, the variance reduces to
\beq {\rm var} (|\tilde\cI(\om)|^2) = 2\pi \<\cI_b\>^4 \cT^2 \ta^2 e^{-n^2 \ta^2 \om^2/2}, \label{back2} \eeq which is exponentially small at the frequency of the signal $\om=2\Om$ if $\Om > 1/(n\ta)$, an inequality most easily satisfied when $n$ (the number of orthonormal filters being employed to compute $\cI$) is large.  In other words, when many filters are used the fluctuation in the background is negligibly small (and we have already shown in the main text that the mean background itself is also small).

To calculate the error in the signal itself, {\it once it is established by the above analysis to be many standard deviations above background}, one notes that the variance of the signal is dominated by the first term of \eqref{back2} while the signal itself by the first term of \eqref{lines}.  Thus the ratio of the signal power to the standard deviation (standard error) in the power, or the first term of \eqref{lines} divided by the square root of the first term of \eqref{back2}, is of order $\cT/\ta \gg 1$, and is independent of $n$.

%\end{appendix}

%$$\<\left(I+I'\right)^2\>-\<I+I'\>^2=2\<I\>^2+2\<I'\>^2+2\left[\<I\>\<I'\>+\ \sum_{j}a_jb_j^* \sum_{k}a_k^*b_k\right]$$
%
%$$-\left(\<I\>^2+\<I'\>^2+2\<I\>\<I'\>\right)$$

%$$ V_1(t) = \sum_{j=1} a_j~e^{i(\om_j t + \phi_j)}$$
%
%$$ V_2(t) = \sum_{j=1}b_j~e^{i(\om_j t + \phi_j)}$$
%
%
%$$ \<I_1\> = \sum_j |a_j|^2;~{\rm and}~\<I_2\> = \sum_k |b_k|^2$$
%
%
%$$ \fr{\si_{I_1}^2}{\bar{I_1}^2} =  \fr{\si_{I_2}^2}{\bar{I_2}^2}= 1$$
%
%$$\cI = I_1 + I_2$$
%
%$$\bf{\cI =V_1V_1^*+V_2V_2^*+\cdots+V_nV_n^*}$$
%
%$$\bf{V_1,V_2,\cdots,V_n}$$
%
%
%$$\sigma_{\cI}^2/\bar{\cI}^2$$

\end{document}